\documentclass[footinbib,aip,amsmath,amssymb,
reprint, onecolumn]{revtex4-1}

\usepackage[utf8]{inputenc}
\usepackage{fontenc}
\usepackage{mathtools}
\usepackage{xpatch}
\usepackage{mathptmx}
\usepackage{hhline}
\usepackage{tikz-cd}
\usepackage{tikz}
\usepackage{tikz-3dplot}
\usepackage{xcolor}
\usepackage{extarrows}
\usepackage{color}
\usepackage{multirow}
\usepackage{adjustbox}
\usepackage{amssymb}
\usepackage{amsmath,bm}
\usepackage{physics}
\usepackage[amsthm, thmmarks, amsmath]{ntheorem}
\usepackage[english]{babel}
\usepackage{amscd}
\definecolor{darkgreen}{RGB}{33,134,115}
\definecolor{darkred}{RGB}{163.2,8.4,46.8}
\usepackage[colorlinks,linktocpage=true,
urlcolor = blue,
linkcolor = darkred,
citecolor = darkgreen
]{hyperref}
\hypersetup{
colorlinks = true
}
\usepackage{lmodern}
\usepackage{caption}
\usepackage{subcaption}
\usepackage{mwe}
\usepackage{wasysym, stackengine, makebox}
\usepackage{float}
\usepackage{dsfont}
\usepackage{newtxmath}
\usepackage{xparse,aliascnt,bookmark}
\usepackage{mathdots}
\usepackage[noabbrev]{cleveref}
\usepackage{comment}
\usepackage{romannum}

\renewcommand{\l}{\left}
\renewcommand{\r}{\right}

\newcommand{\R}{\mathbb{R}}
\newcommand{\Z}{\mathbb{Z}}

\newcommand{\rmi}{\mathrm{i}}

\newcommand{\C}{\mathbb{C}}
\newcommand{\Q}{\mathbb{H}}

\newcommand{\p}{\partial}

\newcommand{\Hi}{\mathcal{H}}

\newcommand{\ex}{\mathrm{e}}
\newcommand{\Ki}{\mathcal{K}}
\newcommand{\Id}{\mathrm{Id}}
\newcommand{\N}{\mathbb{N}}

\renewcommand{\O}{\mathrm{O}}

\newcommand{\B}{\mathcal B}

\DeclareMathOperator{\U}{\mathrm{U}}
\DeclareMathOperator{\SO}{\mathrm{SO}}
\DeclareMathOperator{\SU}{\mathrm{SU}}
\DeclareMathOperator{\nul}{\mathrm nullity}

\DeclareMathOperator{\Sp}{\mathrm Sp}

\newtheorem{theorem}{Theorem}[section]
\newtheorem{corollary}[theorem]{Corollary}
\newtheorem{lemma}[theorem]{Lemma}

\newcommand\isom{\mathrel{\stackon[-0.1ex]{\makebox*{\scalebox{1.08}{\AC}}{=\hfill\llap{=}}}{{\AC}}}}

\newcommand\nvisom{\rotatebox[origin=cc] {-90}{$ \isom $}}

\draft 

\begin{document}
\renewcommand{\thepage}{\arabic{page}}


\title{Classically Frustrated Magnets, Symmetries and
$\Z_2$-Equivariant Topology} 



\author{Shayan Zahedi}
\email[]{zahedi@thp.uni-koeln.de}
\affiliation{Institut für Theoretische Physik, Universität zu Köln, 50937 Cologne, Germany}

\date{\today}

\begin{abstract}

A novel result in $\Z_2$-equivariant homotopy theory is stated, proven, and applied to the topological classification of classically frustrated magnets in the presence of canonical time-reversal symmetry. This result generalizes a lemma that had been key to the homotopical derivation of the renowned Bott-Kitaev periodic table for topological insulators and superconductors. The methods used in the classification of topological insulators and superconductors are here generalized and their generalizations applied to systems that are not quantum mechanical. We distinguish between three symmetry classes  $\mathrm{A\Romannum{3}}$, $\mathrm{BD\Romannum{1}}$, and $\mathrm{C\Romannum{2}}$ depending on the existence and type of canonical time-reversal symmetry. 
For each of these classes, the relevant objects to classify are $\Z_2$-equivariant maps into a Stiefel manifold. The topological classification is illustrated through examples of classically frustrated spin models and is compared to that of Roychowdhury and Lawler (RL). \end{abstract}


\maketitle 

\section{Introduction}

\begin{center}
    \begin{figure}[H]
    \centering
    \adjustbox{scale=.8}{
    \begin{minipage}{.5\textwidth}
            \begin{subfigure}{1.1\textwidth}
                \includegraphics[width=\textwidth]{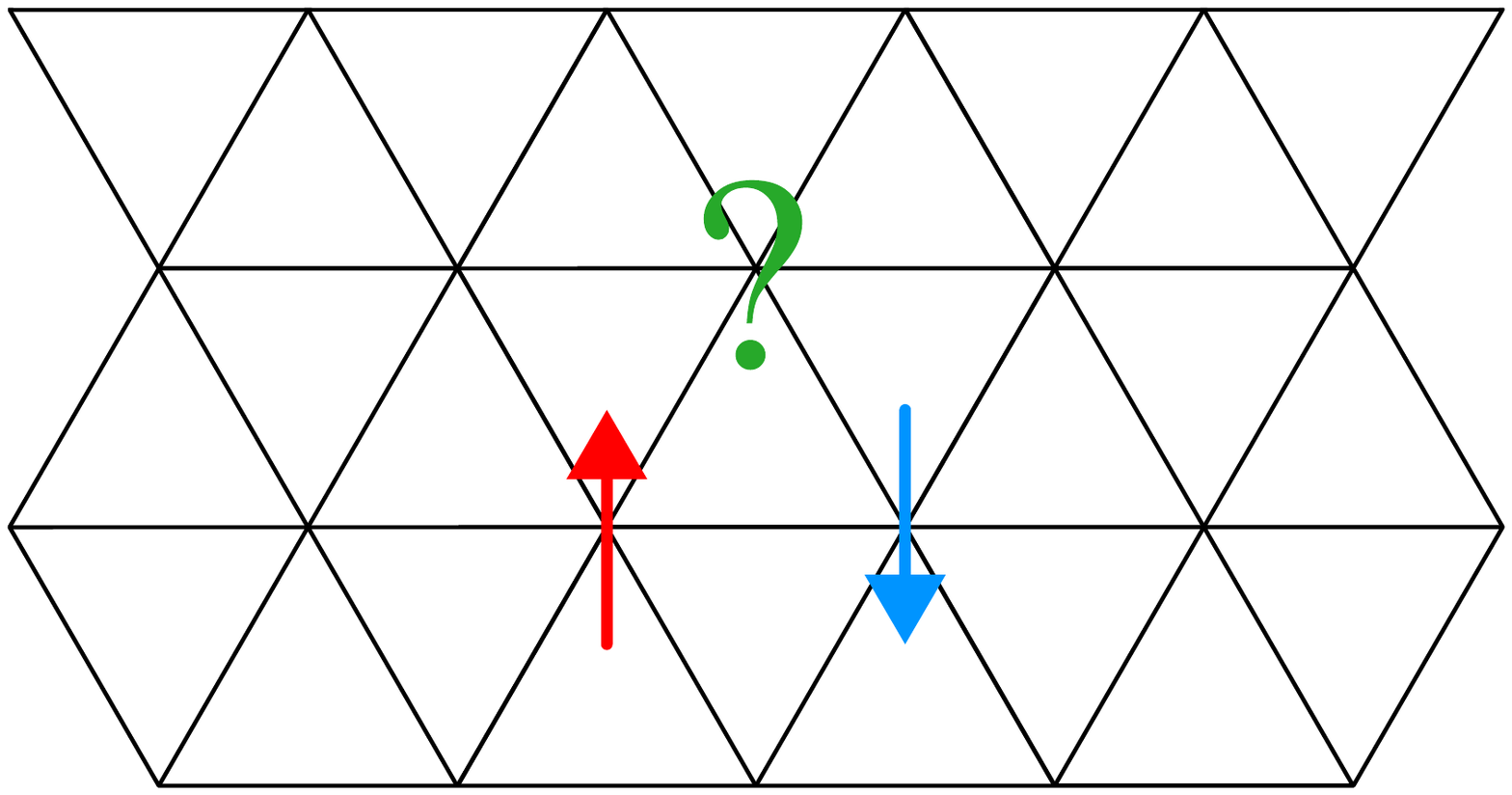}
                \caption{}
                \label{fig:Frustration on triangular lattice}
            \end{subfigure}
        \end{minipage}
        \quad\quad\quad
    \begin{minipage}{.4\textwidth}
        \begin{subfigure}{.65\textwidth}
            \includegraphics[width=\textwidth]{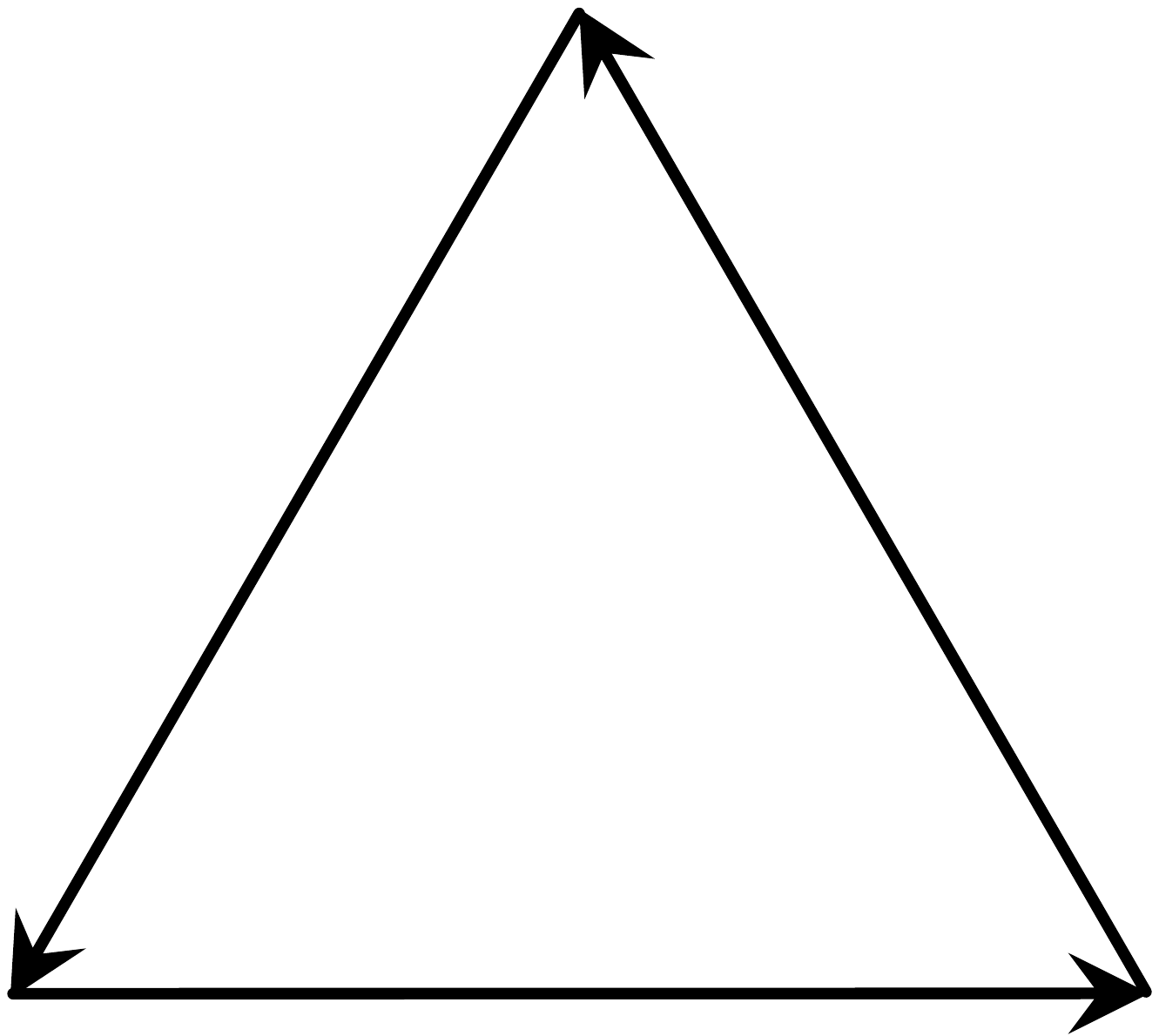}
            \caption{}
            \label{fig:Ground states of triangular antiferromagnet}
        \end{subfigure} \\
        \begin{subfigure}{.65\textwidth}
            \includegraphics[width=\textwidth]{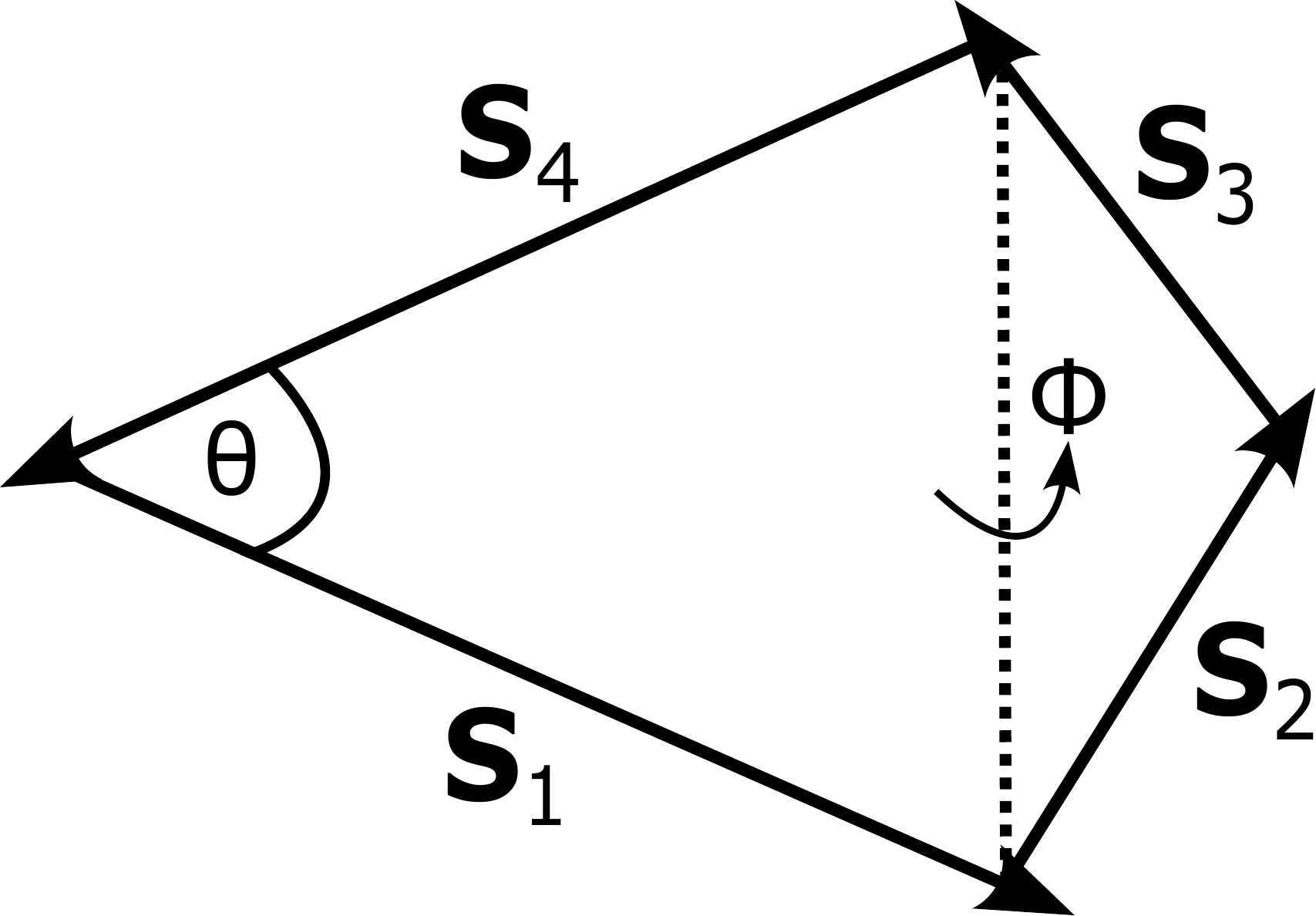}
            \caption{}
            \label{fig:Ground states of pyrochlore Heisenberg antiferromagnet}
        \end{subfigure} 
    \end{minipage}}
    \caption{On a triangular lattice \subref{fig:Frustration on triangular lattice} with antiferromagnetic interactions between nearest neighbors, a configuration in which each spin can be antialigned with all its neighbors is impossible. In other words, the system is frustrated. The sum of spins must add up to zero in each ground state of a cluster of Heisenberg spins. A cluster of three spins \subref{fig:Ground states of triangular antiferromagnet} forms a unique structure, whereas a cluster of four spins \subref{fig:Ground states of pyrochlore Heisenberg antiferromagnet} forms a family of degenerate ground states, parameterized by the structure's two degrees of freedom $\theta$ and $\phi$.}
\end{figure}
\end{center}

The notion of frustration describes the situation where a spin (or several spins) in a spin model cannot find an orientation to fully minimize all the interaction energies with its neighboring spins simultaneously (see \Cref{fig:Frustration on triangular lattice}). In general, frustration is caused either by competing interactions, as in the Villain model \cite{Villain_1977}, or by the lattice structure, as in the triangular, face-centered cubic (fcc) and hexagonal-close-packed (hcp) lattices, with antiferromagnetic nearest neighbor exchange interactions \cite{diep2004frustrated}. When the geometry of a lattice precludes the simultaneous minimization of all interactions, one speaks of a geometrically frustrated system \cite{Moessner2006GeometricalF}.

We are interested in studying the topology of zero modes in classically frustrated systems.
A consequence of frustration is an accidental degeneracy of ground states; that is, two different ground states are not generally related by any symmetry operation. Therefore, Hermitian matrices are not of direct use to describe zero modes in frustrated magnets, as frustration cannot be attributed to the symmetries of a Hamiltonian. Instead, for each ground state of a frustrated system, one identifies the key object: a continuous linear transformation from the space of spin wave degrees of freedom into the space of ground state constraints, the \textit{rigidity matrix} \cite{Roy&Lawler}. Ground state constraints are the conditions that have to be satisfied to put the system under inspection into one of its ground states. Rigidity matrices are rectangular matrices, and their kernels are the spaces of zero modes. They describe the topology of zero modes in frustrated magnets. 

For continuous spins, one can estimate the size of ground state degeneracy of frustrated spin models through the \textit{Maxwellian counting argument} \cite{Moessner2006GeometricalF,Roy&Lawler,lacroix2011introduction}. The key idea is to reorganize the terms in the spin Hamiltonian into constraints, following which the naive degeneracy estimate $\nu$, the \textit{Maxwell counting index}, is obtained. It is the number of ground state degrees of freedom per unit cell and is given as the difference between the total number $N$ of spin wave degrees of freedom per unit cell and the number $M$ of ground state constraints per unit cell, that is, $\nu = N-M$.

\vspace{-.8cm}

\begin{figure}[H]
    \centering
    \includegraphics[width=0.5\textwidth]{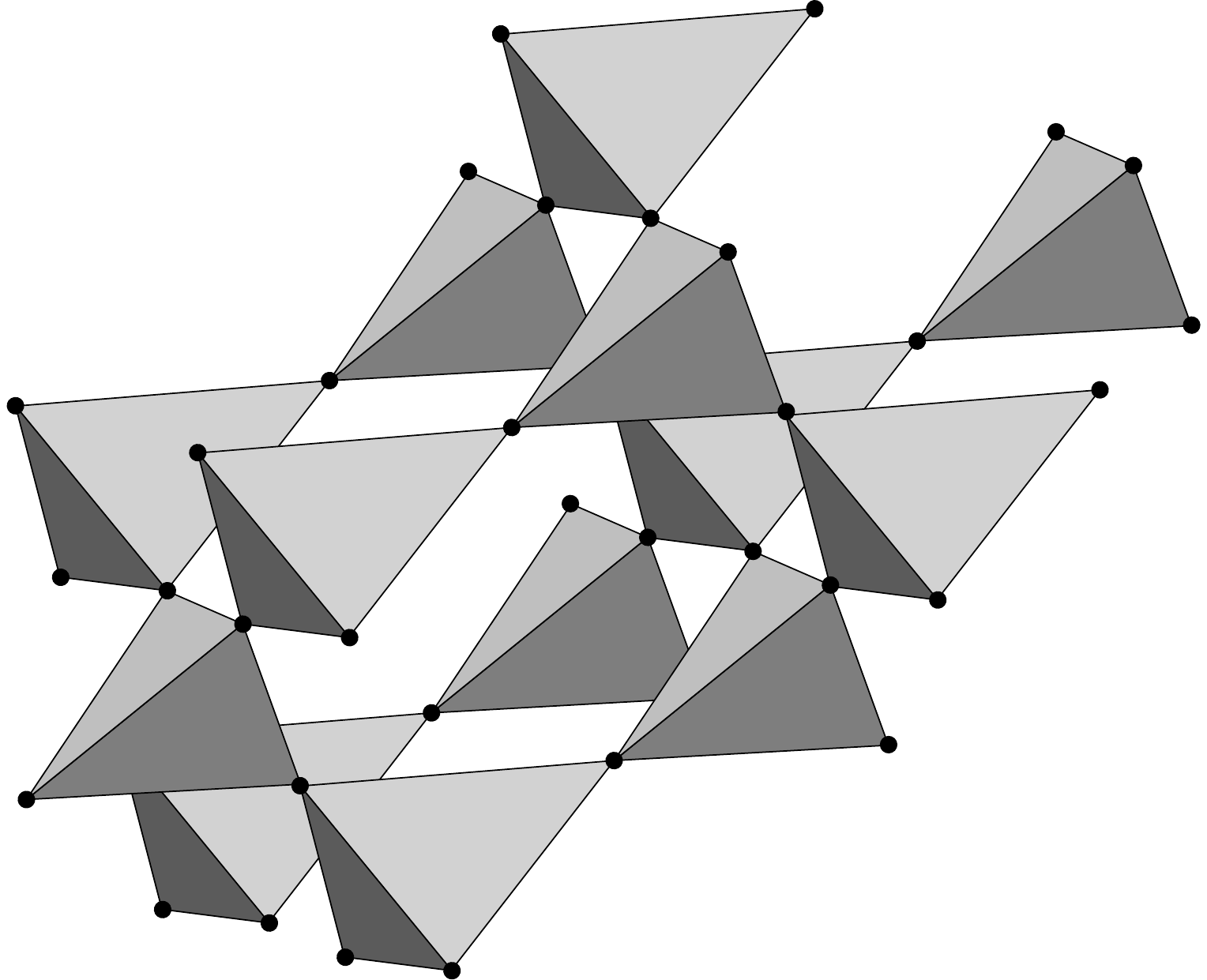}
    \caption{The pyrochlore lattice is a network of vertex-sharing tetrahedra.}
    \label{Fig: The pyrochlore lattice}
\end{figure}

Maxwell introduced such counting to discuss the stability of mechanical systems of joined rods in 1864 \cite{Maxwell}. Chalker and Moessner applied it to frustrated spin systems in 1998 \cite{Moessner_1998}, in particular to the pyrochlore (see \Cref{Fig: The pyrochlore lattice}) Heisenberg antiferromagnet (HAF) \cite{Moessner1998}. It can be shown that the number of ground state degrees of freedom of the entire pyrochlore HAF is extensive as it equals the number of tetrahedra \cite{Reimers1992,Moessner_1998,lacroix2011introduction} and one finds $\nu = 2$ for the pyrochlore HAF \cite[9]{lacroix2011introduction} (see \Cref{fig:Ground states of pyrochlore Heisenberg antiferromagnet}), $\nu = 0$ for the kagome HAF \cite{Roy&Lawler} (see \Cref{fig:Ground states of triangular antiferromagnet}) and $\nu = 1$ for the HAF on a checkerboard lattice \cite{Lieb1999,Canals2002}. The corresponding constraints in the spin Hamiltonians are that the total spin vanishes in each tetrahedron for the pyrochlore HAF, in each triangle for the kagome HAF, and in each checkerboard for the HAF on the checkerboard lattice, respectively \cite{Roy&Lawler}.
 
There is a gap condition for systems described by a rigidity matrix: The number of nonzero singular values is the rank of the rigidity matrix, and the introduction of a new zero mode, i.e., a gap closure, means the reduction of this rank. One makes this gap condition more visible by flattening the singular values of rigidity matrices \cite{Roy&Lawler}. This flattening of singular values is mathematically realized through a strong deformation retraction. Singular value flattened rigidity matrices take values in complex Stiefel manifolds. The corresponding linearized Hamiltonian governing the dynamics of spin waves is given as a bilinear form in terms of the rigidity matrix. 

RL \cite{Roy&Lawler} topologically classify classically frustrated systems in the presence of time-reversal symmetry and $C_2$ rotation symmetry leading to a trivial $\Z_2$-action on the Brillouin torus $T^d = \R^d/2\pi\Z^d$. We develop a topological classification of frustrated systems, resulting in the \Cref{Subtable:By real structures,Subtable:By quaternionic structures,tab:Homotopy groups of complex Stiefel manifolds}, by considering momentum-inversion on the Brillouin torus $T^d$ coming from the sole presence of canonical time-reversal symmetry. Homotopical methods from the classification of topological insulators and superconductors\cite{Kennedy_2015} are generalized (leading to \Cref{eq:Main Lemma}) and applied to systems that are not quantum.
 Depending on the presence and type of canonical time-reversal symmetry, we distinguish between the three symmetry classes $\mathrm{A\Romannum 3}$, $\mathrm{BD\Romannum{1}}$ and $\mathrm{C\Romannum{2}}$. 
We also propose the notion and distinction between ``strong'' and ``weak'' topological invariants. Our consideration is mainly based on strong topological invariants by substituting $T^d$ with the $d$-sphere $S^d$ which reveals sets of homotopy classes beyond those in the Bott-Kitaev periodic
table for topological insulators and superconductors \cite{Kitaev2009,KennedyZ_2015,ludwig,Ryu_2010,Ludwig2008}. The homotopical classification is performed as a function of the number of ground state degrees of freedom per unit cell $\nu$, the underlying lattice dimension $d$, and depends on the realization of canonical time-reversal symmetry. The absence of any further crystalline symmetries is considered in distinction to RL \cite{Roy&Lawler}.

To achieve such a topological classification of frustrated systems in the presence of canonical time-reversal symmetry, we prove and apply one of our main results, \Cref{homotopy groups of equivariant loop space}. \Cref{homotopy groups of equivariant loop space} constitutes a generalization of a result in $\Z_2$-equivariant homotopy theory and establishes an isomorphism between homotopy groups of $\Z_2$-equivariant iterated loop spaces and relative homotopy groups of pairs of iterated loop spaces involving a dimensional shift. More on $\Z_2$-equivariant loop spaces is presented in \cite{gitler2006recent}.

A variant of \Cref{homotopy groups of equivariant loop space} with weaker consequences was previously formulated only for the loop space of some Riemannian symmetric spaces in the context of free-fermion ground states of gapped systems with symmetries \cite{Kennedy_2015} and is here generalized to the iteration of the loop space construction for any $\Z_2$-space. Another weaker variant for the study of three-dimensional insulators with inversion symmetry was intuitively stated in \cite{Turner_2012}. Motivated by RL\cite{Roy&Lawler}, we apply \Cref{homotopy groups of equivariant loop space}
to classify the topology of zero modes in frustrated magnets in the presence of canonical time-reversal symmetry.  
Technically, this theorem always finds applications as long as $\Z_2$-equivariance conditions are present, the $\Z_2$-action on the Brillouin torus $T^d$ is realized through momentum-inversion \cite{das2023realizing,Kennedy_2015,Fiorenza_2016,Hughes_2011,Kaufmann_2016} and $T^d$ can be replaced by $S^d$ at the expense of losing weak topological invariants.

In recent studies, classical spin liquids are classified based on their energy spectrum \cite{fang2023classification,yan2023classification}. A more detailed development and comprehensive exposition of the classification theory with numerous examples can be found in \cite{yan2023classificationDetailed}.

This paper is organized as follows: in chapter \ref{Chapter: The mathematical framework} 
we introduce the physical framework to describe spin waves of frustrated magnets through examples and a mathematical model. The examples include an antiferromagnetic Heisenberg spin chain, a classical HAF on a square lattice with anisotropic next-nearest neighbor exchange interactions, the classical $J_1-J_2$ HAF on a square lattice, and the classical pyrochlore HAF.
In these examples, rigidity matrices are computed, and symmetries are identified. 

In chapter \ref{Chapter:The Topological Classification}, we formulate and prove one of the main results, \Cref{homotopy groups of equivariant loop space}. It is applied to obtain a topological classification of zero modes characterized by time-reversal symmetric (i.e. $\Z_2$-equivariant) rigidity matrices (see \Cref{Subtable:By real structures,Subtable:By quaternionic structures}). Furthermore, it is argued that target spaces of rigidity matrices are complex Stiefel manifolds after minimal assumptions and an appropriate deformation retraction. The homotopical classification is exemplified through the examples in chapter \ref{Chapter: The mathematical framework} and compared to the topological classification of RL \cite{Roy&Lawler} and their time-reversal related symmetry considerations.

\section{The Physical Framework}\label{Chapter: The mathematical framework}
This chapter aims to introduce a physical framework in which we describe spin waves, the linearised degrees of freedom in the ground states of a frustrated system, and ground state constraints. 

We demonstrate physical examples of calculating rigidity matrices and identifying their symmetries. We consider the classical antiferromagnetic Heisenberg spin chain, the classical HAF on a square lattice with anisotropic next-nearest neighbor exchange interactions, the $J_1-J_2$ HAF on a square lattice, and the classical pyrochlore HAF.

The notion of time-reversal symmetry is introduced, and three different $\Z_2$-equivariance conditions are obtained as a consequence.

\subsection{Examples}\label{section:Examples}
Suppose a classically frustrated system on a lattice $\Lambda$ with $\vb0\in\Lambda$ and denote by $\tilde\Lambda$ the subset of positions of magnetic unit cells. Furthermore, suppose we fix one specific ground state configuration in this frustrated system. This ground state is not unique (since frustrated systems have many ground states); e.g., in the examples, the N\'eel ordered state is selected and it can be rotated. But for the following, it is fixed and has a certain periodicity structure for a periodic system.  To count the number of spin degrees of freedom per magnetic unit cell on two-dimensional lattices, we follow the procedure of RL\cite{Roy&Lawler}. In a magnetic unit cell, we count the minimal number of vertices, the minimal number of edges, and a minimal number of faces that are bounded by edges so that, upon translation, the entire system is reconstructed without overlapping. Now, the number of vertices is precisely the number of spin degrees of freedom per unit cell. This procedure of obtaining magnetic unit cells and spin degrees of freedom is generalized to higher dimensional lattices. We declare $0$-cells to be vertices, $1$-cells to be edges, $2$-cells to be faces, and $3$-cells to be volumes. 

Define \begin{equation}\label{spin parametrization}
    f(q,p)\coloneqq \l(\sqrt{1-p^2}\cos(q),\sqrt{1-p^2}\sin(q),p\r).
\end{equation}

Now suppose $k$ $0$-cells and $s$ sites in each unit cell in $\Lambda$ (note that in general $k\leq s$ as described above). We model classical spins as maps $S\colon\Lambda\to S^2$, $\vb x\mapsto S_{\vb x}$. In our fixed ground state configuration and in each unit cell, there exist $q_1,\ldots,q_k,p_1,\ldots,p_k$, and $\vb c_l,d_m\in\Lambda$ with $l\in\{1,\ldots,k\}$ and $m\in\{k+1,\ldots,s\}$ so that $S_{\vb x+\vb c_l} = f(q_l,p_l)$ and $S_{\vb x+\vb d_m} = f(q_{i_m},p_{i_m})$ with $i_m\in\{1,\ldots,k\}$ for all $x\in\tilde\Lambda$. By convention, $\vb c_1 = \vb0$. Suppose there are $r\in\N$ vector-valued constraints per unit cell. In the following examples and throughout this paper, we restrict our attention to spin Hamiltonians $H$ of the form\cite{Roy&Lawler}
\begin{equation}\label{generally rewritten}
    H = \sum_{\vb x\in\tilde\Lambda}\sum_{i=1}^r\l(\overline S^{(i)}_{\vb x}\r)^2
\end{equation}
with \begin{equation}\label{full constraints}
    \overline S^{(i)}_{\vb x}\coloneqq\sum_{l=1}^k\lambda_l^{(i)}S_{\vb x+\vb c_l} + \sum_{m = k+1}^s\mu_m^{(i)}S_{\vb x+\vb d_m}
\end{equation}
denoting the vector-valued constraints and $\lambda_l^{(i)},\mu_m^{(i)}\in\R$. That is, $\overline S^{(i)}_{\vb x} = 0$ for all $i\in\{1,\ldots, r\}$ and $\vb x\in\tilde\Lambda$ characterizes our fixed ground state. We now define the rigidity matrix and assume for brevity $\mu_m^{(i)} = 0$. Expanding \cref{full constraints} up to linear order and denoting by $Df(q_l,p_l)$ the Jacobian matrix of $f$ at $(q_l,p_l)$, the rigidity matrix in position space is obtained (the most general case is obtained by adding $\delta_{\vb d_m}(\vb y)\mu_m^{(i)}Df(q_{i_m},p_{i_m})$ in the right entries of the matrix on the right-hand side of \cref{precise definition of rigidity matrix}, depending on the values of $i_m\in\{1,\ldots,k\}$) by permuting the columns of the matrix on the right-hand side of \cref{precise definition of rigidity matrix} as in\cite{Roy&Lawler}

\begin{equation}\label{precise definition of rigidity matrix}
    r(\vb y)\begin{pmatrix}
    q'_1-q_1\\\vdots\\q'_k-q_k\\p'_1-p_1\\\vdots\\p'_k-p_k
    \end{pmatrix}\coloneqq\begin{pmatrix}
        \delta_{\vb0}(\vb y)\lambda_1^{(1)}Df(q_1,p_1) & \cdots & \delta_{\vb c_k}(\vb y)\lambda_k^{(1)}Df(q_k,p_k)\\\vdots & & \vdots\\
        \delta_{\vb0}(\vb y)\lambda_1^{(r)}Df(q_1,p_1) & \cdots & \delta_{\vb c_k}(\vb y)\lambda_k^{(r)}Df(q_k,p_k)
    \end{pmatrix}\begin{pmatrix}
        q'_1-q_1\\p'_1-p_1\\\vdots\\q'_k-q_k\\p'_k-p_k
    \end{pmatrix},
\end{equation}
with \begin{equation}
    \delta_{\vb x}(\vb y)\coloneqq\begin{cases}
        1 & \text{if }\vb x = \vb y\\
        0 & \text{otherwise.}
    \end{cases}
\end{equation}
The variables $q'_1,\ldots,q'_k,p'_1,\ldots,p'_k$ denote the linearized ground state degrees of freedom. 
Regarding the later discussion in Chapter \ref{general model building}, the linearized Hamiltonian governing the spin-wave dynamics is thus characterized by $H = R^\dagger R$ in which the rigidity matrix $r$ is the position space representation of the \textit{rigidity operator} $R$ (see \cref{position space representation}). The position space representation of $H$ hence reads \begin{equation}
    h(\vb y)=\sum_{\vb x\in\Lambda}r(\vb x-\vb y)^\dagger r(\vb x).
\end{equation}

Thus, the following approach to deriving a rigidity matrix for a fixed ground state configuration of a frustrated spin system is used: One starts from a spin Hamiltonian that can be rewritten into a sum of non-negative terms as in \cref{generally rewritten}. In the fixed ground state configuration, the non-negative terms must vanish in each magnetic unit cell.  
These ground state constraints are expanded to linear order around a chosen ground state configuration. The rigidity matrix $r\colon\Z^d\to\l(\C^{M\times N}\r)^{\Z_2}$ in position space is now constructed as follows. All elements of the rigidity matrix are precisely the coefficients of all first-order terms in the spin wave expansion of the ground state constraints (up to a permutation of columns as in \cref{precise definition of rigidity matrix}). This approach is schematically illustrated by RL\cite{Roychowdhury_2018} and more rigorously exemplified in the following. 

We first illustrate the approach with the classical antiferromagnetic Heisenberg spin chain having the spin Hamiltonian (realizing spins as functions $S\colon\Z\to S^2$, $x\mapsto S_x$)
\begin{equation}\label{new_spin_chain}
    H = J\sum_{x\in\Z}S_{x}S_{x+1} = \frac{J}{2}\sum_{x\in\Z}\l(S_x+S_{x+1}\r)^2+\mathrm{const.}
\end{equation}
with positive antiferromagnetic coupling constant $J$. Ground state constraints are therefore 
\begin{equation}\label{gs_constraint_chain}
    S_x+S_{x+1}=0
\end{equation}
for all $x\in\Z$, telling us that the ground state is a N\'eel ordered state, which was expected. We fix the N\'eel axis. The N\'eel axis, as also in the following examples, is not unique since it can be rotated. But for the following, we assume it to be fixed. We parameterize the spins via \cref{spin parametrization} and linearize the ground state constraints in \cref{gs_constraint_chain} around the fixed N\'eel axis to receive, as in \cref{precise definition of rigidity matrix}, the following rigidity matrix in position space. 

\begin{equation}
    r(y) = \begin{pmatrix}
        \delta_0(y)-\delta_1(y) & 0\\0 & \delta_0(y)+\delta_1(y)
    \end{pmatrix}.
\end{equation}
Following the definition of the rigidity matrix in \cref{precise definition of rigidity matrix} closely, one would initially object and point out that one should expect a $3\times 2$ matrix. However, the discarded row is a zero row and should thus be eliminated\cite{Roy&Lawler}. In the following topological classification, we consider rigidity matrices whose singular values could assume zeros at isolated momenta and not be trivial over the entire Brillouin torus. The topological information of a rigidity matrix is encoded in its non-trivial entries. 
With the Fourier transformation in \cref{Best fourier}, the rigidity matrix 
\begin{equation}\label{rigidity chain}
    \tilde r(k) = \begin{pmatrix}
        1-e^{\rmi k} & 0\\0&1+e^{\rmi k}
    \end{pmatrix}
\end{equation}
in momentum space is obtained.
This rigidity matrix is of symmetry class $\mathrm{BDI}$ (see \Cref{Different equivariance conditions}) and exhibits the Maxwell counting index $\nu = 0$, see \Cref{Subtable:By real structures}. The rigidity matrix in \cref{rigidity chain} does not exhibit a $C_2$ rotation symmetry and, therefore, falls beyond the classification of RL.

\begin{figure}[H]
    \centering
    \includegraphics[width=0.5\textwidth]{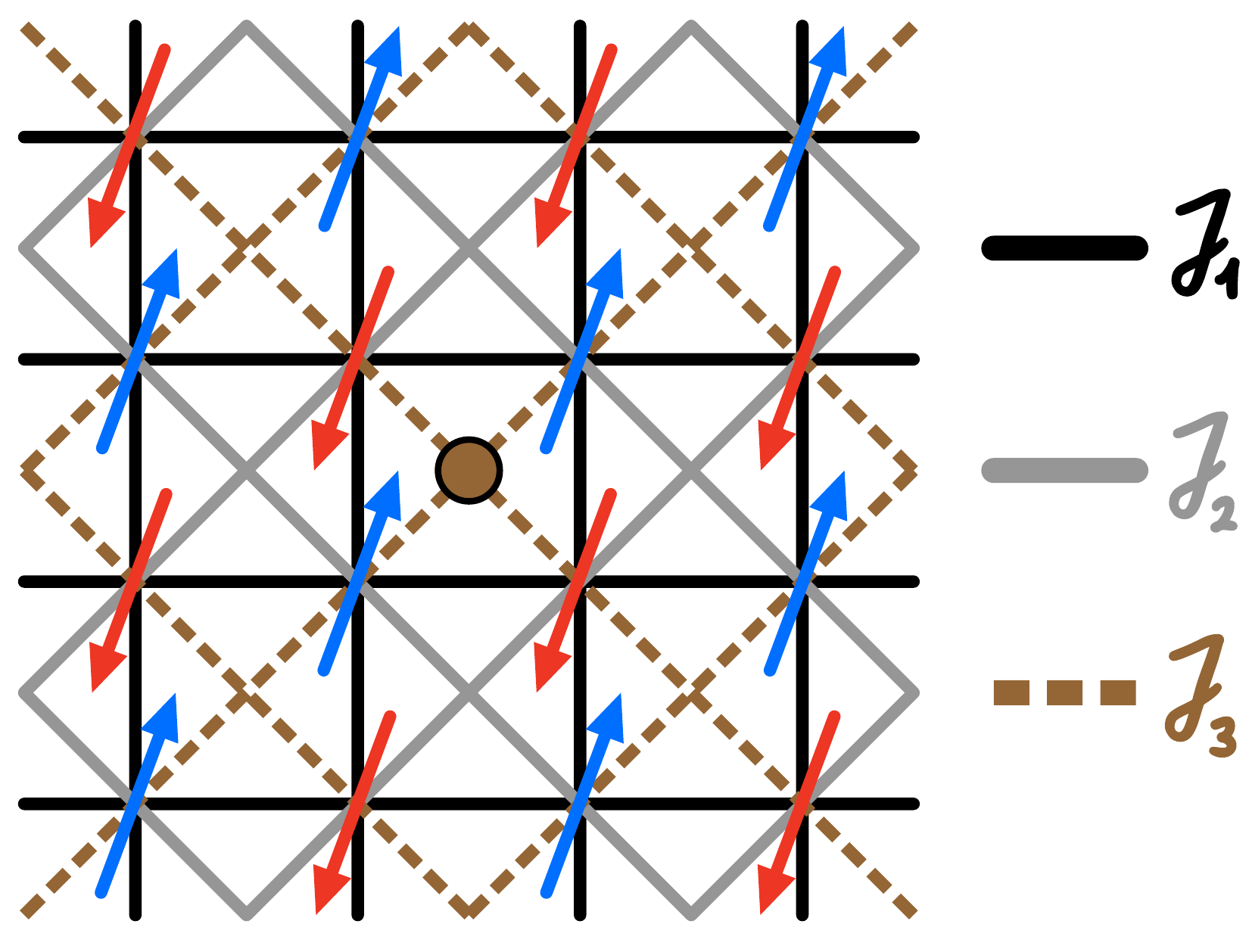}
    \caption{
    The N\'eel state of the HAF on a square lattice with anisotropic next-nearest neighbor exchange interactions represents a ground state. It is symmetric under a global spin flip followed by a $\pi/2$ rotation around the brown dot of the brown dual square lattice. The nearest neighbor exchange interaction $J_1$ and next-nearest neighbor exchange interactions $J_2$ and $J_3$ are also illustrated. 
    }
    \label{Fig:Altermagnetic state}
\end{figure}

Altermagnets display a new type of collinear magnetism distinct from ferromagnetism and conventional antiferromagnetism \cite{Mazin2022,Sinova2022,Sinova2022_2}. The altermagnetic Hubbard model \cite{das2023realizing} inspires us to consider a N\'eel state of the classical HAF on the square lattice with anisotropic antiferromagnetic next-nearest neighbor exchange interactions depicted in \Cref{Fig:Altermagnetic state}.
The corresponding spin Hamiltonian reads (realising spins as functions $S\colon\Z^2\to S^2$, $\vb x\mapsto S_{\vb x}$ and denoting $\vb a_1=(1,0)$ and $\vb a_2=(0,1)$ as primitive vectors) 

\begin{subequations}
\label{eq:Altermagnetic Hamiltonian}
    \begin{align}
    \begin{split}
        H &=J_1\sum_{\vb x\in\Z^2}\l(S_{\vb x}S_{\vb x+\vb a_1} + S_{\vb x}S_{\vb x+\vb a_2}\r)+ J_2\sum_{\vb x\in V}\l(S_{\vb x}S_{\vb x+\vb a_1+\vb a_2}+S_{\vb x+\vb a_1}S_{\vb x+\vb a_2}\r)\\&\quad+J_3\sum_{\vb x\in V}\l(S_{\vb x}S_{\vb x+\vb a_1-\vb a_2}+S_{\vb x-\vb a_2}S_{\vb x+\vb a_1}\r)
    \end{split}\\
    \begin{split}
        &=\frac{J_1}{4}\sum_{\vb x\in V}\l(S_{\vb x}+S_{\vb x+\vb a_1}+S_{\vb x+\vb a_2}+S_{\vb x+\vb a_1+\vb a_2}\r)^2 + \frac{J_1}{4}\sum_{\vb x\in V}\l(S_{\vb x}+S_{\vb x+\vb a_1}+S_{\vb x-\vb a_2}+S_{\vb x+\vb a_1-\vb a_2}\r)^2\\
        &\quad +\frac{J_1}{2}\abs{\frac{J_2}{J_1}-\frac{1}{2}}\sum_{\vb x\in V}\l[\l(S_{\vb x+\vb a_1}\pm S_{\vb x+\vb a_2}\r)^2+\l(S_{\vb x}\pm S_{\vb x+\vb a_1+\vb a_2}\r)^2\r]\\
        &\quad +\frac{J_1}{2}\abs{\frac{J_3}{J_1}-\frac{1}{2}}\sum_{\vb x\in V}\l[\l(S_{\vb x+\vb a_1}\pm S_{\vb x-\vb a_2}\r)^2+\l(S_{\vb x}\pm S_{\vb x+\vb a_1-\vb a_2}\r)^2\r]+\mathrm{const.}\label{eq:Rewriting in terms of constraints}
    \end{split}
    \end{align}
\end{subequations}

with $V\coloneqq\Z(\vb a_1+\vb a_2)\oplus\Z(\vb a_1-\vb a_2)$ denoting the lattice of the magnetic unit cells. In \cref{eq:Rewriting in terms of constraints}, the sign $\pm$ is used in the case $J_1\lessgtr 2J_i$ for $i\in\{2,3\}$. The magnetic unit cell consists of two $0$-cells, four $1$-cells, and two $2$-cells (being doubled in size in comparison to the nuclear unit cell, which is a square plaquette).

In the case of $J_1 = 2J_i$ for both $i=2,3$, the spin Hamiltonian becomes 
\begin{equation}\label{eq:Example J1-J2 HAF}
    H = \frac{J_1}{4}\sum_{\vb x\in\Z^2}\l(S_{\vb x}+S_{\vb x+\vb a_1}+S_{\vb x+\vb a_1}+S_{\vb x+\vb a_1+\vb a_2}\r)^2+\mathrm{const.}
\end{equation}
and the magnetic unit cell coincides with the nuclear unit cell. Ground states are therefore obtained for \begin{equation}\label{gs constraints}
    S_{\vb x}+S_{\vb x+\vb a_1}+S_{\vb x+\vb a_1}+S_{\vb x+\vb a_1+\vb a_2}=0
\end{equation} for all $\vb x\in\Z^2$. Again, parametrizing the spins via \cref{spin parametrization} and linearizing the ground state constraints in \cref{gs constraints} along the fixed positive $x$-axis in the N\'eel ordered state depicted in \Cref{Fig:Altermagnetic state}, we obtain the following rigidity matrix in position space according to \cref{precise definition of rigidity matrix}. 

\begin{equation}
    r(\vb y) = \begin{pmatrix}
        \delta_{\vb 0}(\vb y)-\delta_{\vb a_1}(\vb y)-\delta_{\vb a_2}(\vb y)+\delta_{\vb a_1+\vb a_2}(\vb y) & 0\\0 & \delta_{\vb 0}(\vb y)+\delta_{\vb a_1}(\vb y)+\delta_{\vb a_2}(\vb y)+\delta_{\vb a_1+\vb a_2}(\vb y)
    \end{pmatrix}
\end{equation}
A subsequent Fourier transformation by \cref{Best fourier} reveals (compare this to RL\cite{Roy&Lawler})
\begin{equation}
    \tilde r(\vb k)=\begin{pmatrix}
        1-e^{\rmi k_x}-e^{\rmi k_y}+e^{\rmi(k_x+k_y)} & 0\\
        0 & 1+e^{\rmi k_x}+e^{\rmi k_y}+e^{\rmi(k_x+k_y)}
    \end{pmatrix},\label{eq:Rigidity matrix of RL}
\end{equation}
the rigidity matrix in momentum space. This rigidity matrix is of symmetry class $\mathrm{BD\Romannum1}$ (see \Cref{Different equivariance conditions}) and exhibits the Maxwell counting index $\nu = 0$, see \Cref{Subtable:By real structures}.

Now, in the case $J_1>2J_i$ for both $i\in\{2,3\}$ and $J_2\neq J_3$, we obtain the ground state constraints $S_{\vb x}+S_{\vb x+\vb a_1}+S_{\vb x+\vb a_2}+S_{\vb x+\vb a_1+\vb a_2}=0$, $S_{\vb x}+S_{\vb x+\vb a_1}+S_{\vb x-\vb a_2}+S_{\vb x+\vb a_1-\vb a_2}=0$, $S_{\vb x+\vb a_1}-S_{\vb x+\vb a_2}=0$, $S_{\vb x}-S_{\vb x+\vb a_1+\vb a_2}=0$, $S_{\vb x+\vb a_1}-S_{\vb x-\vb a_2}=0$ and $S_{\vb x}-S_{\vb x+\vb a_1-\vb a_2}=0$ for all $\vb x\in V$ and the rigidity matrix reads \begin{equation}\label{eq:Rigidity matrix for anisotropic nnn interactions}
        \tilde r(\vb k) = \begin{pmatrix}
            1+e^{\rmi k_x} & -1-e^{-\rmi k_y} & 0 & 0\\
            1+e^{\rmi k_y} & -1-e^{-\rmi k_x} & 0 & 0\\
            0 & e^{-\rmi k_y}-1 & 0 & 0\\
            1-e^{\rmi k_x} & 0 & 0&0\\
            0&1-e^{-\rmi k_x}&0&0\\
            e^{\rmi k_y}-1&0&0&0\\
            0&0&1+e^{\rmi k_x}&1+e^{-\rmi k_y}\\
            0&0&1+e^{\rmi k_y}&1+e^{-\rmi k_x}\\
            0&0&0&1-e^{-\rmi k_y}\\
            0&0&1-e^{\rmi k_x}&0\\
            0&0&0&e^{-\rmi k_x}-1\\
            0&0&e^{\rmi k_y}-1&0
        \end{pmatrix}
    \end{equation}
being once again of symmetry class $\mathrm{BD\Romannum1}$ and displaying two submatrices with individual Maxwell counting index $\nu = 4$, see \Cref{Subtable:By real structures}. It should be stated that over-constrained systems (systems with $M>N$) can also exhibit frustration. In fact, in the search for highly frustrated magnets ($\nu>0$), many frustrated magnets with $\nu<0$ have been found. The frustration in such magnets is rather a consequence of the presence of competing exchange interactions between the spins rather than of the geometry of the underlying lattice\cite{Roy&Lawler}, as is illustrated in \Cref{Fig:Altermagnetic state}.

Squaring the $\pi/2$-rotation symmetry, we find the following additional $\Z_2$-equivariance condition $\tilde r(-\vb k) = [\l((-\sigma_1)\oplus(\sigma_1\otimes\sigma_1)\r)\oplus(\sigma_1\oplus(-\sigma_1\otimes\sigma_1))]\tilde r(\vb k)[\sigma_1\oplus\sigma_1]$. The rigidity matrix in \cref{eq:Rigidity matrix for anisotropic nnn interactions} does not exhibit a $C_2$ rotation symmetry and is thus not classifiable by the methods of RL. 

Another example is the classical pyrochlore HAF.

\begin{figure}[H] 
\captionsetup[subfloat]{margin=10pt,format=hang}
    \begin{subfigure}[b]{0.5\linewidth}
    \centering
    \adjustbox{scale=.9}{\includegraphics[scale = 0.35]{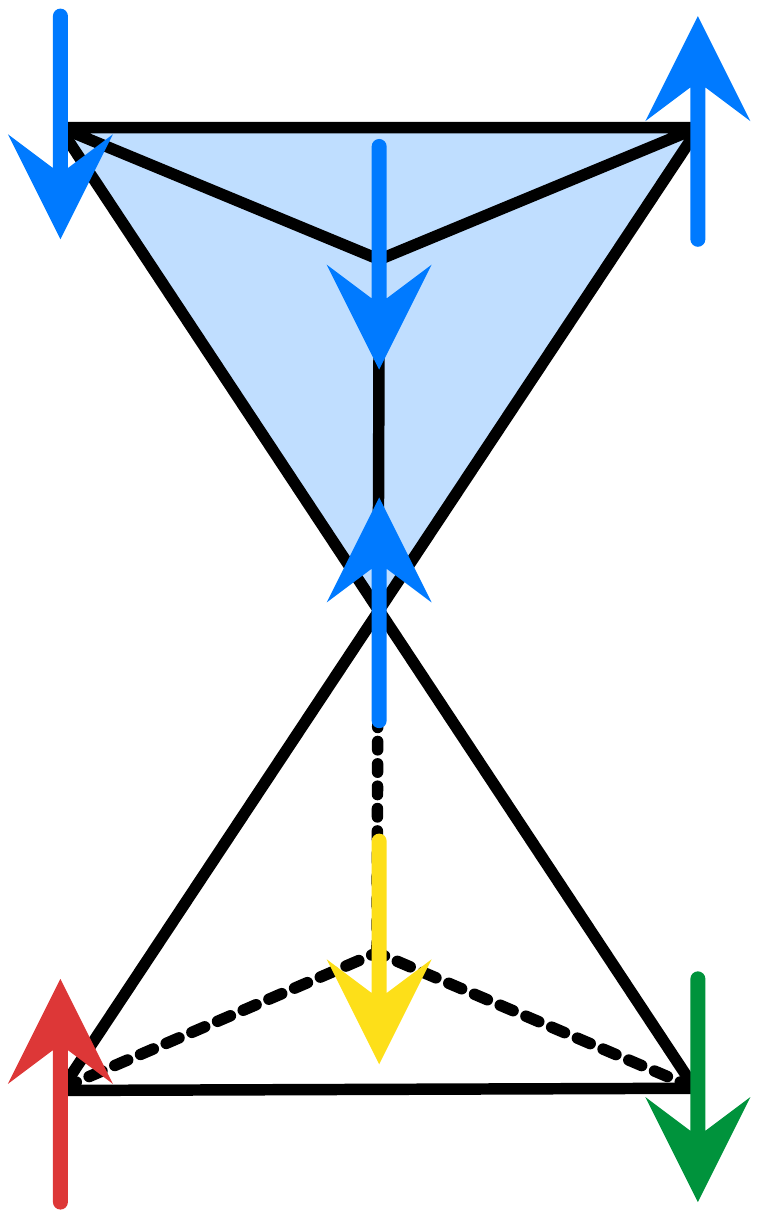}}
    \vspace{-1.5cm}
    \caption{A collinear ground state} 
    \label{subfig:Pyrochlore GS} 
  \end{subfigure}
  \begin{subfigure}[b]{0.5\linewidth}
    \centering
    \adjustbox{scale=.9}{\includegraphics[scale = 0.35]{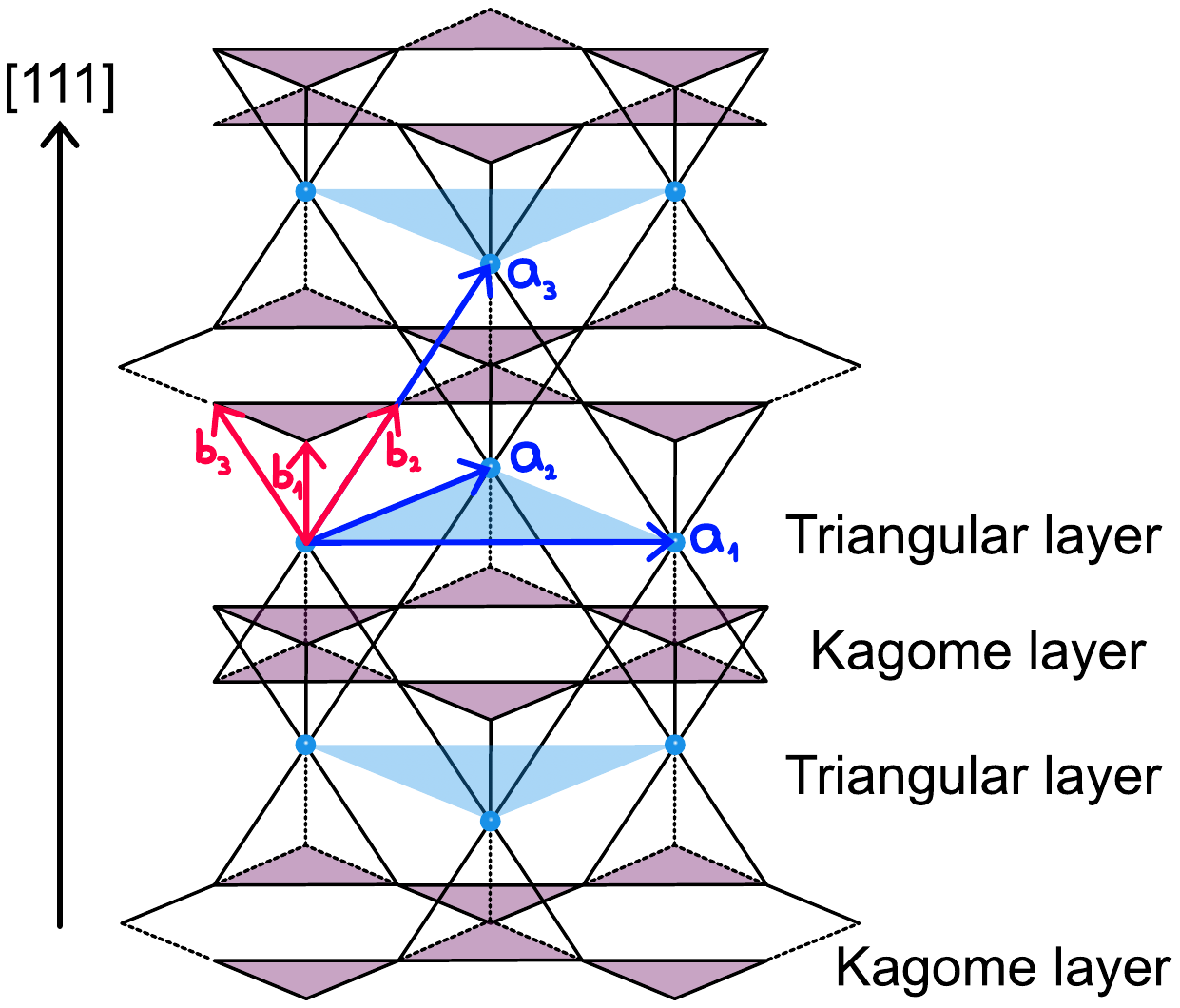}}
    \vspace{-1.5cm}
    \caption{The pyrochlore lattice along the $[111]$ direction} 
    \label{subfig:pyrochlore lattice [111] direction} 
  \end{subfigure}
  \caption{A collinear ground state configuration for the pyrochlore HAF is depicted in \subref{subfig:Pyrochlore GS}. The spins sitting on the vertices of the blue tetrahedron belong to the same unit cell, whereas the yellow, red, and green spins belong to neighboring unit cells, respectively. \subref{subfig:pyrochlore lattice [111] direction} The pyrochlore lattice is an alternating stacking of kagome (purple) and triangular (blue) layers along the $[111]$, body diagonal, direction. The primitive vectors $\vb a_1$, $\vb a_2$ and $\vb a_3$ span the lattice for the unit cells and the internal vectors $\vb b_1 \coloneqq(\vb a_3-\vb a_2)/2$, $\vb b_2\coloneqq\vb a_3/2$ and $\vb b_3\coloneqq(\vb a_3-\vb a_1)/2$ describe the positions of spins on the kagome layers.}
  \label{Pyrochlore example} 
\end{figure}

There are two types of tetrahedra in the pyrochlore lattice, both depicted in \Cref{subfig:Pyrochlore GS}.
The unit cell of the pyrochlore lattice consists of four $0$-cells (indicated by the four blue spin axes in \Cref{subfig:Pyrochlore GS}), twelve $1$-cells (all the links between the vertices of the corner-sharing tetrahedra in \Cref{subfig:Pyrochlore GS}), eight $2$-cells (all the faces of the corner-sharing tetrahedra in \Cref{subfig:Pyrochlore GS}) and three $3$-cells (the two corner-sharing tetrahedra in \Cref{subfig:Pyrochlore GS} and a neighboring volume). 

Orienting the entire pyrochlore lattice along the $[111]$ direction, as depicted in \Cref{subfig:pyrochlore lattice [111] direction}, one observes that the pyrochlore lattice consists of alternating kagome and triangular layers stacked on top of each other \cite{Tabata2006,lacroix2011introduction}. The lattice describing the positions of the unit cells for the pyrochlore lattice is, therefore, $\mathbb P = \Z\vb a_1\oplus\Z\vb a_2\oplus\Z\vb a_3\subseteq\R^3$ with $\vb a_1 = (1,0,0)$, $\vb a_2 = \l(1/2,\sqrt{3}/2,0\r)$ and $\vb a_3\coloneqq\l({3}/8,\sqrt{3}/8,\sqrt{13}/4\r)$. One of the points belonging to this lattice is the corner-sharing point in \Cref{subfig:Pyrochlore GS}.
\vspace{-1cm}

The spin Hamiltonian reads ($J>0$)

\begin{subequations}
\label{eq:Altermagnetic Hamiltonian}
    \begin{align}
    \begin{split}
        H &= J\sum_{\vb x\in\mathbb P}\big[S_{\vb x}S_{\vb x+\vb b_1}+S_{\vb x}S_{\vb x+\vb b_2}+S_{\vb x}S_{\vb x+\vb b_3}+S_{\vb x+\vb b_1}S_{\vb x+\vb b_2}+S_{\vb x+\vb b_2}S_{\vb x+\vb b_3}+S_{\vb x+\vb b_3}S_{\vb x+\vb b_1}\\&\quad+S_{\vb x}S_{\vb x-\vb b_1}+S_{\vb x}S_{\vb x-\vb b_2}+S_{\vb x}S_{\vb x-\vb b_3}+S_{\vb x-\vb b_1}S_{\vb x-\vb b_2}+S_{\vb x-\vb b_2}S_{\vb x-\vb b_3}+S_{\vb x-\vb b_3}S_{\vb x-\vb b_1}\big]
    \end{split}\\
    \begin{split}
        &=\frac{J}{2}\sum_{\vb x\in\mathbb P}\l[L_{1,\vb x}^2+L_{2,\vb x}^2\r]+\mathrm{const.}\label{eq:Exampel Pyrochlore HAF}
    \end{split}
    \end{align}
\end{subequations}
with $L_{1,\vb x}\coloneqq S_{\vb x}+S_{\vb x+\vb b_1}+S_{\vb x+\vb b_2}+S_{\vb x+\vb b_3}$ and $L_{2,\vb x}\coloneqq S_{\vb x}+S_{\vb x-\vb b_1}+S_{\vb x-\vb b_2}+S_{\vb x-\vb b_3}$ for all $\vb x\in\mathbb P$. Ground states of the pyrochlore HAF are obtained for $L_{2,\vb x} = 0 = L_{1,\vb x}$ for all $\vb x\in\mathbb P$. The simplex lattice (describing the positions of the constraints) is a diamond lattice \cite{Roychowdhury2022SupersymmetryOT}. Considering a fixed collinear ground state \cite{Moessner_1998,lacroix2011introduction} of the pyrochlore HAF and, without loss of generality, the direction of the collinear order along the positive $x$-axis, we obtain the rigidity matrix

\begin{equation}\label{eq:Rigidity matrix for pyrochlore heisenberg antiferromagnet}
    \tilde r(\vb k) = \begin{pmatrix}
    1 & -1&1&-1&0&0&0&0\\
    1&-e^{\rmi(k_2-k_3)}&e^{-\rmi k_3}&-e^{\rmi(k_1-k_3)}&0&0&0&0\\
    0&0&0&0&1&1&1&1\\
    0&0&0&0&1&e^{\rmi(k_2-k_3)}&e^{-\rmi k_3}&e^{\rmi(k_1-k_3)}
\end{pmatrix}.
\end{equation}

It is of symmetry class $\mathrm{BD\Romannum{1}}$ (see \Cref{Different equivariance conditions}).  Here, the rigidity matrix $\tilde r$ consists of two independent blocks representing $\nu = 2$ systems, see \Cref{Subtable:By real structures}. Since the pyrochlore lattice is three-dimensional, the rigidity matrix in \cref{eq:Rigidity matrix for pyrochlore heisenberg antiferromagnet} does not exhibit a $C_2$ inversion symmetry and thus falls beyond the classification of RL.

\subsection{A Mathematical Model}\label{general model building}

\begin{figure}[H]
    \centering
    \includegraphics[width=0.7\textwidth]{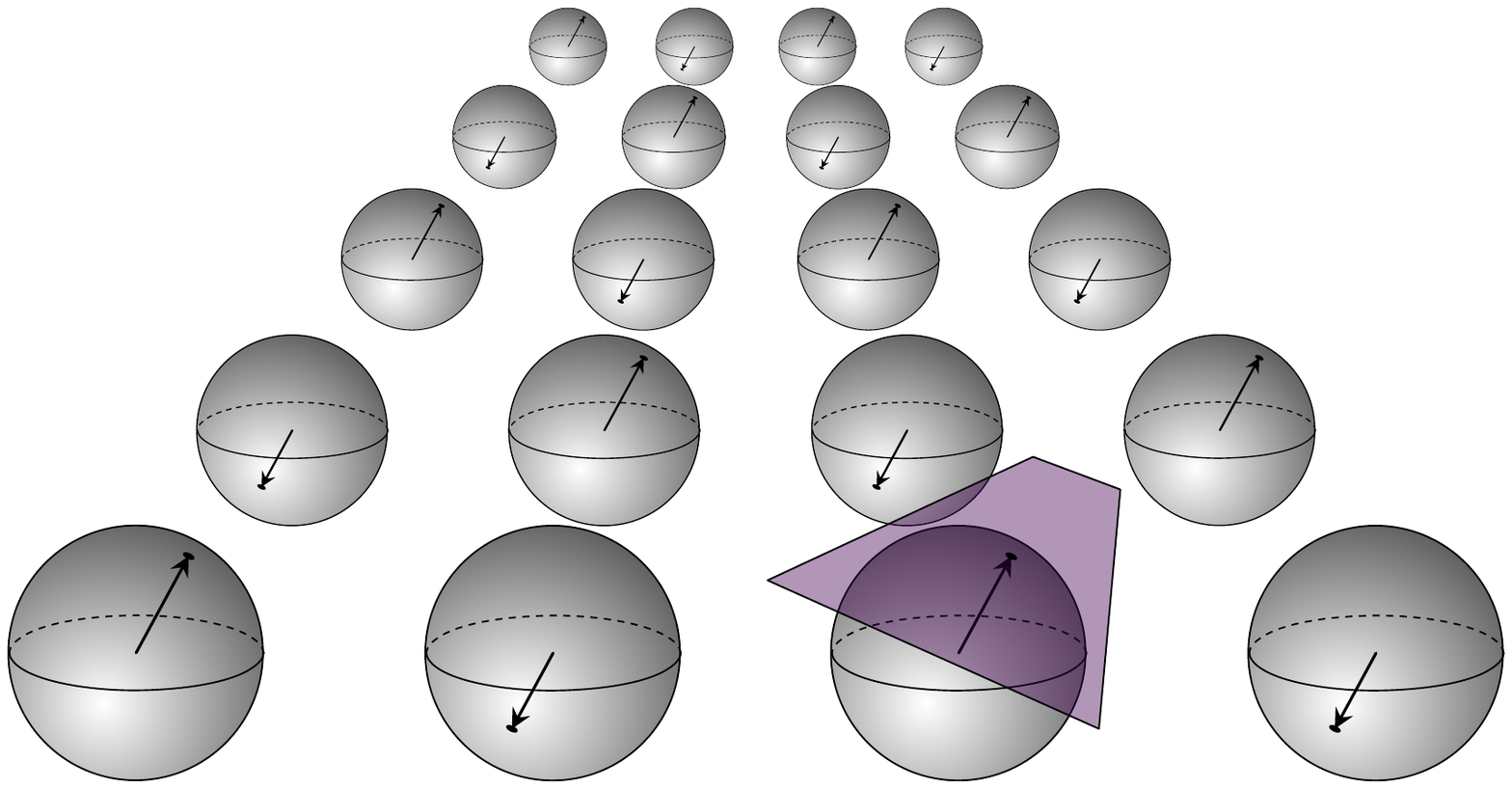}
    \caption{A N\'eel ordered state serves as one of many ground states of the $J_1-J_2$ HAF on a square lattice \cite{Roy&Lawler}. The N\'eel vector can be chosen arbitrarily, tracing out the whole $2$-sphere $S^2$ (grey), and expanding around any ground state amounts to considering linearised degrees of freedom coming from the planes (purple) orthogonal to the fixed spin axes (schematic illustration).}
    \label{Fig:Square lattice explaining Hilbert space}
\end{figure}

We consider a $d$-dimensional underlying lattice $\Z^d$ (position space) with minimal distance normalized to $1$ and associate with each lattice position a space $\C^N$ of $N$ linearly independent spin wave degrees of freedom in a frustrated system. In the case of three-component spins having unit magnitude, each spin can be associated with a $2$-sphere $S^2$ and is restricted to a subspace of $S^2$ in the ground state. Choosing a particular ground state configuration (which amounts to fixing spin axes), each spin's linearised degrees of freedom live in the plane orthogonal to the spin axes, see \Cref{Fig:Square lattice explaining Hilbert space}.

In analogy to phase space, we assume our number of linearised degrees of freedom to be even. Therefore, the planes perpendicular to the spin axes (being isomorphic to one another) are modeled by $\C^N\cong\R^{2N}$. The complex Hilbert space containing the spin waves is hence modeled by 

\begin{equation}
    \Hi_d^N\coloneqq\ell^2\l(\Z^d,\C^N\r)=\left\{\varphi\colon\Z^d\to\C^N\biggm\vert\sum_{i=1}^N\sum_{\vb x\in\Z^d}\abs{\varphi_i(\vb x)}^2<\infty\right\}
\end{equation}
with the scalar product 

\begin{equation}
    \l\langle\varphi,\psi\r\rangle_{\Hi_d^N}\coloneqq \sum_{\vb x\in\Z^d}\left\langle\varphi(\vb x),\psi(\vb x)\right\rangle_{\C^N}
\end{equation}

which serves as a conventional tool for upcoming calculations. The square summability condition on our spin waves is needed to perform a Fourier transformation later.

The ground state constraints of a magnetically frustrated spin system constitute a system of equations per magnetic unit cell across the entire lattice, which define the structure of ground state configurations. 
Although the form of the lattice $\mathbb G$ describing the positions of the constraints is model-dependent because it depends, among other things, on the interactions between the spins, we consider $\mathbb G=\Z^d$ without loss of generality. The reason is that an isomorphism $\mathbb G\cong\Z^d$ is already sufficient to remain in the framework of the upcoming topological classification. 

Moreover, upon considering the ground state constraints up to linear order in a spin wave expansion, we will obtain $M$ linearly independent ground state constraints so that the appropriate space modeling ground state constraints in spin wave expansions becomes $\Hi_d^M$. It will later serve as the target space of so-called \textit{rigidity operators} which mediate between spin wave degrees of freedom and ground state constraints in spin wave expansions.

We immediately see that $\B_d^N\coloneqq\{\delta_{\vb x}e_i\mid\vb x\in\Z^d\text{ and }1\leq i\leq N\}$ constitutes an orthonormal basis for $\Hi_d^N$, where $(e_i)_{i=1}^N$ constitutes the standard basis for $\C^N$.

For every $\vb a\in\Z^d$, we define translation operators as the unitary operators $t_{\vb a}\colon\Hi_d^N\to\Hi_d^N$, $t_{\vb a}(\delta_{\vb x}e_i)\coloneqq\delta_{\vb x+\vb a}e_i$, and linear extension. We realize time-reversal operators as either real or quaternionic structures\cite{Alldridge_2019}, i.e., as antiunitary operators $T = UK\colon\Hi_d^N\to\Hi_d^N$ squaring either to $+\Id_{\Hi_d^N}$ or to $-\Id_{\Hi_d^N}$. Here, $U$ is a unitary operator, and $K$ is the operation of complex conjugation concerning the basis $\mathcal B_d^N$. Linearised degrees of freedom and ground state constraints assume values in $\R^{2N}\cong\C^N$ and $\R^{2M}\cong\C^M$. These spaces are essentially classical phase spaces. This is the reason for realizing time-reversal operators on both spaces $\Hi_d^N$ and $\Hi_d^M$ as real or quaternionic structures (having the effect of reversing the sign of the symplectic structure of phase space).

Linear operators $R\colon\Hi_d^N\to\Hi_d^M$ acting as\footnote{In \cref{position space representation}, $(e_j)_{j=1}^N$ on the left-hand side and $(e_i)_{i=1}^M$ on the right-hand side constitute the standard bases for $\C^N$ and $\C^M$, respectively.} \begin{equation}\label{position space representation}
R(\delta_{\vb y}e_j) = \sum_{i=1}^M\sum_{\vb x\in\Z^d}\tilde R_{ij}(\vb x,\vb y)\delta_{\vb x}e_i,
\end{equation}
are uniquely characterised by their matrix representation $\tilde R\colon\Z^d\times\Z^d\to\C^{M\times N}$. We specifically classify the cases in which $\tilde U(\vb x,\vb y) = \delta_{\vb x}(\vb y)J$ with $J$ being in its \textit{standard form}\footnote{Note that in the case $T^2 = -\Id_{\Hi_d^N}$ one can show easily that $N$ must be even. The matrix $I_N$ denotes the $N\times N$ identity matrix. The $\sigma_i$ denote Pauli matrices.} \cite{Freeman2004,Thompson1988NormalFF,Roy&Lawler,dresselhaus2007group}
\begin{equation}\label{Realisation of time reversal operator}
    J = \begin{cases}
        I_N&\text{for }T^2 = +\Id_{\Hi_d^N},\\
        I_{N/2}\otimes\rmi\sigma_2&\text{for }T^2 = -\Id_{\Hi_d^N}.
    \end{cases}
\end{equation}
Motivated by the examples in chapter \ref{section:Examples}, we will work with translation invariant time-reversal symmetric \textit{rigidity operators} $R\colon\Hi_d^N\to\Hi_d^M$. That is, they satisfy $t_{\vb a}R = Rt_{\vb a}$\footnote{We use the same notation for translation operators on both $\Hi_d^N$ and $\Hi_d^M$. Similarly, this is done for the notation of upcoming Fourier transformations.}, and $T_2R = RT_1$, or equivalently $\tilde R(\vb x+\vb a,\vb y+\vb a) = \tilde R(\vb x,\vb y)$ and 
$J_2\overline{\tilde R(\vb x,\vb y)}J_1^\dagger = \tilde R(\vb x,\vb y)$ for all $\vb a,\vb x,\vb y\in\Z^d$ implying that $R$ is uniquely characterised by $r\colon\Z^d\to\l(\C^{M\times N}\r)^{\Z_2}$, $r(\vb x)\coloneqq R(\vb x,\vb 0)$. Here, $T_1$ and $T_2$ are both either real or quaternionic structures on $\Hi_d^N$ and $\Hi_d^M$ respectively and \begin{equation}\label{eq:Z_2-fixed point set of C^{MxN}}
    \l(\C^{M\times N}\r)^{\Z_2}\cong\begin{cases}\C^{M\times N} & \text{no time-reversal symmetry},\\
        \R^{M\times N} & \text{for }T_1^2 = +\Id_{\Hi_d^N}\text{ and }T_2^2 = +\Id_{\Hi_d^M},\\
        \Q^{M/2\times N/2}&\text{for }T_1^2 = -\Id_{\Hi_d^N}\text{ and }T_2^2 = -\Id_{\Hi_d^M},
    \end{cases}
\end{equation}
is the $\Z_2$-fixed point set concerning the $\Z_2$-action $g\Lambda\coloneqq J_2\overline\Lambda J_1^\dagger$, where $g\in\Z_2$ denotes the nontrivial element. 

Through the Fourier transformation $F\colon\Hi_d^N\to L^2\l(T^d,\C^N\r)\eqqcolon\Ki_d^N$, given by $F(\delta_{\vb x}e_i)\coloneqq\ex_{\vb x}e_i$, with $\ex_{\vb x}(\vb k)\coloneqq e^{\rmi\vb k\vb x}/(2\pi)^{d/2}$, and linear extension, we see that rigidity operators $R'\coloneqq FRF^\dagger\colon\allowbreak\Ki_d^N\to\Ki_d^M$ act diagonally on momentum space because $R'\tilde\varphi = \tilde r\tilde\varphi$ with 
\begin{equation}\label{Best fourier}
    \tilde r(\vb k)\coloneqq\sum_{\vb x\in\Z^d}r(\vb x)e^{\rmi\vb k\vb x}.
\end{equation} 

That is, $R'$ acts through multiplication by the \textit{rigidity matrix} $\tilde r\colon T^d\to\C^{M\times N}$. Simply using the rank-nullity theorem \cite{friedberg2003linear}, we express the Maxwell counting index in terms of rigidity matrices as $\nu = \nul\tilde r-\nul\tilde r^\dagger$, constituting a special case of a so-called analytical index \cite{nakahara2003geometry}.

The corresponding linearized Hamiltonian governing the spin wave dynamics is $H = R^\dagger R$ \cite{Roy&Lawler,Roychowdhury_2018}, i.e., a bilinear form in terms of $R$. It acts diagonally in momentum space through multiplication by $\tilde h = \tilde r^\dagger\tilde r\colon T^d\to\C^{N\times N}$.
All zero modes in a frustrated model can be explained in the framework of rigidity operators whose kernel $\ker H = \ker R$ contains the zero modes. Any zero mode $\tilde\varphi\in F\ker R$ equivalently satisfies $\tilde\varphi(\vb k)\in\ker\tilde r(\vb k)$ for almost all $\vb k\in T^d$. Therefore, a homotopical classification of rigidity matrices directly addresses how frustration can be preserved by perturbations \cite{Roy&Lawler}.

We are specifically concerned with rigidity operators $R$ whose rigidity matrices $\tilde r$ are continuous \cite{Roy&Lawler,Roychowdhury_2018} and $\Z_2$-equivariant, distinguishing the following three cases. Compared to RL\cite{Roy&Lawler}, our classification considers $\Z_2$-equivariance.

\begin{table}[H]
        \centering
        \begin{tabular}{c|c|c}
           Label & Time-reversal symmetry  &  $\Z_2$-equivariance of $\tilde r\colon T^d\to\C^{M\times N}$\\\hline
            $\mathrm{AIII}$&no & trivial $\Z_2$-equivariance\\
            $\mathrm{BDI}$&yes, $T_1^2 = +\Id_{\Hi_d^N}$, $T_2^2 = +\Id_{\Hi_d^M}$ & $\tilde r\l(-\vb k\r) = \overline{\tilde r\l(\vb k\r)}$\\
            $\mathrm{CII}$ & yes, $T_1^2 = -\Id_{\Hi_d^N}$, $T_2^2 = -\Id_{\Hi_d^M}$ & $\tilde r\l(-\vb k\r) = \l(I_{M/2}\otimes\sigma_2\r)\overline{\tilde r(\vb k)}\l(I_{N/2}\otimes\sigma_2\r)$
        \end{tabular}
        \caption{The $\Z_2$-equivariance conditions on the rigidity matrix $\tilde r\colon T^d\to\C^{M\times N}$ as a consequence of the existence and type of canonical time-reversal symmetry.}
        \label{Different equivariance conditions}
    \end{table} The $\Z_2$-action of time-reversal on the Brillouin torus $T^d=\R^d/2\pi\Z^d\cong (I/\p I)^d$, with $I\coloneqq[-\pi,\pi]$, is given as $g[\vb k] = [-\vb k]$.
    Time-reversal invariant momenta live in the finitely generated subgroup $(T^d)^{\Z_2} = \langle[\pi e_1],\ldots,[\pi e_d]\rangle$ of order $\abs{(T^d)^{\Z_2}} = 2^d$.

\section{The Topological Classification}\label{Chapter:The Topological Classification}

The examples in chapter \ref{section:Examples} motivate us to define the space of rigidity matrices $R_{dM}^N$ as the subspace of the space of continuous base point preserving ${\Z_2}$-equivariant maps $\mathrm{Map}_*\l(T^d,\C^{M\times N}\r)^{\Z_2}$ with the following additional properties. The elements of $R_{dM}^N$ have maximum rank everywhere. They are such that a linear interpolation between an element of $\mathrm{Map}_*\l(T^d,\C^{M\times N}\r)^{\Z_2}$ and the corresponding matrix in which all singular values have been replaced by $1$ constitutes a $\Z_2$-homotopy. This is a nontrivial assumption since the unitary matrices entering a singular value decomposition of a rigidity matrix may not necessarily be $\Z_2$-equivariant by themselves. These properties are reflected by the examples in chapter \ref{section:Examples}. Although our examples in \ref{section:Examples} are more specific in the sense that already the unitary matrices entering the singular value decomposition of the rigidity matrices in \cref{rigidity chain,eq:Rigidity matrix of RL,eq:Rigidity matrix for anisotropic nnn interactions,eq:Rigidity matrix for pyrochlore heisenberg antiferromagnet} exhibit $\Z_2$-equivariance, we adopt this slightly more general viewpoint that $\Z_2$-equivariance of the product of the unitary matrices with the matrix of singular values is preserved under continuous deformation to the rigidity matrix with all singular values replaced by $1$.

Rank-constancy and maximality implement the assumption that the ground state constraints are linearly independent. A topological classification for the general case of fluctuating ranks can be reduced to the constant-and-maximum-rank case\footnote{\label{Footnote:Rank-reducing rigidity matrices}Indeed, after performing the flattening technique of singular values for the general case, the target spaces of rigidity matrices become unions of Stiefel manifolds and upon choosing a base point, one selects a single Stiefel manifold.}. The base point condition on rigidity matrices is merely a matter of selecting bases in $\C^N$ and $\C^M$, respectively, at a specific momentum. 

Linearly interpolating between the positive singular values of the elements of $R_{dM}^N$ and $1$ gives a strong deformation retraction. In the case $M\geq N$, the strong deformation retract is precisely $\mathrm{Map}_*\l(T^d,V_N\l(\C^M\r)\r)^{\Z_2}$, where \begin{equation}
    V_N\l(\C^M\r) =\l\{A\in\C^{M\times N}\mid A^\dagger A=I_N\r\}\cong\U(M)/\U(M-N)
\end{equation} denotes the complex Stiefel manifold. In the case $M<N$, the strong deformation retract is homeomorphic to $\mathrm{Map}_*\l(T^d,V_n\l(\C^m\r)\r)^{\Z_2}$, where $m\coloneqq\max(M,N)$ and $n\coloneqq\min(M,N)$, and thus exhibits isomorphic homotopy groups. A $\Z_2$-homeomorphism is given by pointwise transposition. The $\Z_2$-action on $V_n(\C^m)$ is precisely the one described in \Cref{Different equivariance conditions}.
Now, modding out $\Z_2$-homotopy, one is left with the set of homotopy classes $\l[T^d,V_n(\C^m)\r]_*^{\Z_2}$ of base point preserving and $\Z_2$-equivariant maps $T^d\to V_n(\C^m)$ containing strong and weak topological invariants. These sets of homotopy classes classify zero modes in frustrated systems in the presence or absence of canonical time-reversal symmetry. 

In the following, the cube $I^d=[-\pi,\pi]^d$ is viewed as a $\Z_2$-space through the nontrivial action $g\vb k\coloneqq-\vb k$ and the iterated loop space $\Omega^{d}X$ of the based $\Z_2$-space $(X,x_0)$ denotes the space of all maps $I^{d}\to X$ which send $\p I^{d}$ to the base point $x_0\in X^{\Z_2}$. The action of $\Z_2$ on $\Omega^{d}X$ is considered to be $(g,f)\mapsto(\vb k\mapsto gf(-\vb k))$.
Replacing the domain $T^d$ (periodic case) with the $d$-sphere $S^d\cong I^d/\p I^d$ (free case), we obtain a topological classification by strong topological invariants at the expense of losing weak topological invariants \cite{Roychowdhury2022SupersymmetryOT,kennedy&guggenheim,Ryu_2010,Schnyder_2009}. Strong and weak topological invariants coincide in dimension $d = 1$ because $T^1 = S^1$. In the following, we are interested in calculating 
\begin{equation}
    \l[S^d,V_n\l(\C^m\r)\r]_*^{\Z_2}\cong\l[\l(I^d,\p I^d\r),(V_n(\C^m),E)\r]_{\Z_2}\cong\pi_0\l(\l(\Omega^dV_n(\C^m)\r)^{\Z_2}\r)
\end{equation}
with $E\coloneqq(e_1\ \cdots\ e_n)$ denoting the canonical $n$-frame.

To unlock a deduction of our sought-after homotopical classification of zero modes in frustrated systems, we formulate one of the main results of this paper. Namely, \Cref{homotopy groups of equivariant loop space} establishes an isomorphism between the homotopy groups of $\Z_2$-equivariant iterated loop spaces and relative homotopy groups of pairs of iterated loop spaces, involving a dimensional shift. This \Cref{homotopy groups of equivariant loop space} is of vital importance because strong topological invariants are contained in sets of path components (constituting the ``zeroth homotopy groups'') of $\Z_2$-equivariant iterated loop spaces of complex Stiefel manifolds.

The following \Cref{homotopy groups of equivariant loop space} generalizes a result that is used for the topological classification of free-fermion ground states of gapped systems with symmetries \cite{Kennedy_2015} from the loop space of certain Riemannian symmetric spaces to the iteration of the loop space construction for any $\Z_2$-space. Another weaker variant was previously stated for the study of three-dimensional insulators with inversion symmetry\cite{Turner_2012} and is now generalized to any dimension of the underlying lattice.
The chosen base points for the left and right-hand side of \cref{eq:Main Lemma} are the constant maps to $x_0$\footnote{We usually suppress the base point in the notation if the context makes the choice of base point clear.}. From a purely mathematical point of view, the following theorem should be compared to the familiar statement $\pi_D\l(\Omega^{d+1}X\r)\cong\pi_{D+d+1}(X)$\cite{dieck2008algebraic_loop}.

\begin{theorem}\label{homotopy groups of equivariant loop space}
    Let $X$ be  a $\Z_2$-space, $x_0\in X$ a $\Z_2$-fixed point and $D,d\geq0$. Then, there is an isomorphism \begin{equation}\label{eq:Main Lemma}      \pi_D\l(\l(\Omega^{d+1}X\r)^{\Z_2}\r)\cong\pi_{D+1}\l(\Omega^dX,\l(\Omega^dX\r)^{\Z_2}\r).
    \end{equation}
    \begin{proof}
        For $D\geq1$, we set $T\coloneqq I^D\times[0,\pi]$ and $J_T\coloneqq\partial I^D\times[0,\pi]\cup I^D\times\{\pi\}$. We set $T=[0,\pi]$ and $J_T=\{\pi\}$ in the case $D=0$. We use the homeomorphism of triples $\phi\colon(T,\partial T,J_T)\to\l(I^{D+1},\partial I^{D+1},J^D\r)$, $(\vb k,p)\mapsto(\vb k,2p-\pi)$, with $\vb k\in I^D$ and $p\in[0,\pi]$, to induce the isomorphism \begin{equation}        \phi^*\colon\pi_{D+1}\l(\Omega^{d}X,\l(\Omega^{d}X\r)^{\Z_2}\r)\xrightarrow{\cong}\l[\l(T,\partial T,J_T\r),\l(\Omega^dX,\l(\Omega^dX\r)^{\Z_2},c_{x_0}\r)\r]\eqqcolon K_{D+1}^d
        \end{equation}
        by pullback $\phi^*([f])\coloneqq[f\circ\phi]$. The bijectivity of $\phi^*$ is because $\phi$ is a homeomorphism. Moreover, the homomorphism property of $\phi^*$ in dimensions $D\geq1$ results from the fact that $\phi$ only affects the last coordinate, but loop concatenation is performed in the first coordinate (see \cref{concatenation}). 
        The base point $c_{x_0}$ denotes the constant map to $x_0$. The inverse of $\phi$ cuts the cube $I^{D+1}$ in half, and the claimed isomorphism in \cref{eq:Main Lemma} says that a $\Z_2$-equivariant map is wholly determined by half of the cube since the other half follows from $\Z_2$-equivariance.
        Now let $[f]\in\pi_{D}\l(\l(\Omega^{d+1}X\r)^{\Z_2}\r)$ and define the map \begin{equation}\label{hinrichtung}
            \tilde f\colon (T,\partial T,J_T)\to\l(\Omega^{d}X,\l(\Omega^{d}X\r)^{\Z_2},\allowbreak c_{x_0}\r) \text{ by } \tilde f(\vb k,p)\coloneqq f(\vb k)(p,-)
        \end{equation} which represents the homotopy class $\l[\tilde f\r]\in K_{D+1}^d$. Indeed, for any $(\vb k,p)\in J_T$ it is clear that $\tilde f(\vb k,p) = c_{x_0}$ as $f\in\Omega^D\l(\l(\Omega^{d+1}X\r)^{\Z_2}\r)$, and the remaining case $\tilde f(\vb k,0)(-\vb q) =f(\vb k)(-(0,\vb q)) = gf(\vb k)(0,\vb q) = g\tilde f(\vb k,0)(\vb q)$ for any $\vb k\in I^D$ and $q\in I^d$ shows that indeed $\tilde f(\p T)\subseteq\l(\Omega^dX\r)^{\Z_2}$.
        This induces the map \begin{equation}
            \eta\colon\pi_D\l(\l(\Omega^{d+1}X\r)^{\Z_2}\r)\to K_{D+1}^d \text{ defined by } \eta[f]\coloneqq\l[\tilde f\r].
        \end{equation}
        The map $\eta$ is well defined because for $[f] = [g]\in\pi_D\l(\l(\Omega^{d+1}X\r)^{\Z_2}\r)$, there is a homotopy $H\colon\big(I^D\times[0,1],\partial I^D\times[0,1]\big)\allowbreak\to\l(\l(\Omega^{d+1}X\r)^{\Z_2},c_{x_0}\r)$ from $H_0 = f$ to $H_1 = g$. The homotopy $H$ gives rise to the homotopy $\tilde H\colon(T\times[0,1],\allowbreak\partial T\times[0,1],J_T\times[0,1])\to\l(\Omega^dX,\l(\Omega^dX\r)^{\Z_2},c_{x_0}\r)$ defined by $\tilde H_t(\vb k,p)\coloneqq H_t(\vb k)(p,-)$ from $\tilde H_0 = \tilde f$ to $\tilde H_1 = \tilde g$ implying $\eta[f] = \eta[g]$.

        The inverse map \begin{equation}
            \eta^{-1}\colon K_{D+1}^d\to\pi_D\l(\l(\Omega^{d+1}X\r)^{\Z_2}\r)\text{ reads } \eta^{-1}[h] = \l[h'\r]
        \end{equation} in which
        $h'\colon\l(I^D,\p I^D\r)\to\l(\l(\Omega^{d+1}X\r)^{\Z_2},c_{x_0}\r)$ is the map \begin{equation}\label{inverse map}
            h'(\vb k)(p,\vb q)\coloneqq\begin{cases}
                gh(\vb k,-p)(-\vb q)&\text{for }p\in[-\pi,0],\\
                h(\vb k,p)(\vb q)&\text{for }p\in[0,\pi].
            \end{cases}
        \end{equation}
        \Cref{inverse map} immediately yields $h'(\vb k)(-(p,\vb q)) = gh'(\vb k)(p,\vb q)$ for any $\vb k \in I^D$ and $(p,\vb q)\in I^{d+1}$, verifying that $h'(\vb k)$ is indeed $\Z_2$-equivariant for all $\vb k\in I^{D}$.
        Furthermore, $\eta^{-1}$ is well defined because for $[h] = [k]\in K_{D+1}^d$ there exists a homotopy
            $F\colon(T\times[0,1],\allowbreak\partial T\times[0,1],J_T\times[0,1])\to\l(\Omega^dX,\l(\Omega^dX\r)^{\Z_2},c_{x_0}\r)$      
         from $F_0 = h$ to $F_1 = k$. The map $G\colon\l(I^D,\p I^d\r)\to\l(\l(\Omega^{d+1}X\r)^{\Z_2},c_{x_0}\r)$ defined by $G_t\coloneqq\l(F_t\r)'$ is a homotopy from $G_0 = h'$ to $G_1 = k'$ and thus, indeed, $\eta^{-1}[h] = \eta^{-1}[k]$. We now prove that  $\eta^{-1}\eta = \Id_{\pi_D\l(\l(\Omega^{d+1}X\r)^{\Z_2}\r)}$ and $\eta\eta^{-1} = \Id_{K_{D+1}^d}$. Combining \cref{hinrichtung,inverse map}, we see that for any $[f]\in\pi_D\l(\l(\Omega^{d+1}X\r)^{\Z_2}\r)$, $\eta^{-1}\eta([f])$ is represented by 
         \begin{equation}
             \begin{aligned}
             &\begin{cases}
                 g\tilde f(\vb k,-p)(-\vb q)&\text{for }p\in[-\pi,0],\\
                 \tilde f(\vb k,p)(\vb q)&\text{for }p\in[0,\pi].
             \end{cases} \\&= f(\vb k)(p,\vb q)
         \end{aligned}
         \end{equation}
         for all $\vb k\in I^D$ and $(p,\vb q)\in I^{d+1}$ because $f\in\Omega^D\l(\l(\Omega^{d+1}X\r)^{\Z_2}\r)$. Hence, $\eta^{-1}\eta([f]) = [f]$. Similarly, choosing any $[h]\in K_{D+1}^d$, a representative of $\eta\eta^{-1}([h])$ is, by \cref{hinrichtung}, the restriction of \cref{inverse map} to $p\in[0,\pi]$, thence $\eta\eta^{-1}([h])=[h]$. 
        
        As for now, we have proven that $\eta$ is a bijection for all $D\geq 0$. Now consider the case $D\geq 1$ and let $[f],[g]\in\pi_D\l(\l(\Omega^{d+1}X\r)^{\Z_2}\r)$. We easily find 
        \begin{equation}\label{concatenation}
        \begin{aligned}
            \widetilde{f+g}(\vb k,p)&=\begin{cases}
                f(2k_1+\pi,k_2,\ldots,k_D)(p,-)&\text{for }k_1\in[-\pi,0],\\
                g(2k_1-\pi,k_2,\ldots,k_D)(p,-)&\text{for }k_1\in[0,\pi].
            \end{cases}
            \\&=\begin{cases}
                \tilde f(2k_1+\pi,k_2,\ldots,k_D,p)&\text{for }k_1\in[-\pi,0],\\
                \tilde g(2k_1-\pi,k_2,\ldots,k_D,p)&\text{for }k_1\in[0,\pi].
            \end{cases}
            \\&=\tilde f + \tilde g(\vb k,p)
        \end{aligned}
        \end{equation}
        for all $(\vb k,p)\in T$ which implies $\eta\l([f]+[g]\r) = \eta([\widetilde{f+g}]) = \eta([\tilde f]+[\tilde g])= \eta[f] + \eta[g]$ and shows that $\eta$ is also a homomorphism for $D\geq1$.
    \end{proof}
\end{theorem}

Applying \Cref{homotopy groups of equivariant loop space} to the special case $d = 0$, immediately provides

\begin{corollary}\label{special case of equivariant homotopy groups}
    Let $X$ be a $\Z_2$-space, $x_0\in X$ a $\Z_2$-fixed point and $D\geq0$. There is an isomorphism \begin{equation}
        \pi_D\l(\l(\Omega X\r)^{\Z_2}\r)\cong\pi_{D+1}\l(X,X^{\Z_2}\r).
    \end{equation}\hfill$\square$
\end{corollary}

Through the algorithm in \cref{eq:The Algorithm}, we will prove inductively that ($\nu = N-M$)
\begin{equation}\label{trivial regime}
    \l[\l(I^d,\p I^d\r),\l(V_{m-\abs{\nu}}(\C^m),E\r)\r]_{\Z_2} = 0 \text{ for all } \abs{\nu}\geq\lceil d/2\rceil
\end{equation}
in the presence of canonical time-reversal symmetry. Moreover, we will derive the nontrivial result
\begin{equation}
    \l[\l(I,\p I\r),\l(\U(m),I_m\r)\r]_{\Z_2}\cong\Z
\end{equation}
that holds for both symmetriy classes $\mathrm{BDI}$ and $\mathrm{CII}$.
Furthermore, the following \Cref{tab:Topological invariants with Time Reversal Symmetry} summarises the sets of homotopy classes $\big[\big(I^d,\p I^d\big),\allowbreak\big(V_{m-\abs{\nu}}\allowbreak(\C^m),E\big)\big]_{\Z_2}$ containing strong topological invariants in the presence of canonical time-reversal symmetry up to the Brillouin torus dimension $d = 3$.

\begin{table}[H]
\begin{subtable}[c]{0.5\textwidth}
\centering
   \begin{tabular}{|c|c|c|c|}\hline
         \multirow{2}{*}{$\abs{\nu}$} &\multicolumn{3}{|c|}{$d$}\\\cline{2-4}
            &  $1$ & $2$ & $3$\\\hhline{|=|=|=|=|}
          $0$ & $\mathbb Z$ & $\star$ & $\star$ \\
          $1$ & $0$ & $0$ & $\star$\\
          $\geq2$ & $0$ & $0$ & $0$\\\hline
    \end{tabular}
    \subcaption{Symmetry class $\mathrm{BD\Romannum 1}$}
    \label{Subtable:By real structures}
    \end{subtable}
\begin{subtable}[c]{0.5\textwidth}
\centering
   \begin{tabular}{|c|c|c|c|}\hline
         \multirow{2}{*}{$\abs{\nu}$} &\multicolumn{3}{|c|}{$d$}\\\cline{2-4}
            &  $1$ & $2$ & $3$ \\\hhline{|=|=|=|=|}
          $0$ & $\mathbb Z$ & $\star$ & $\star$\\
          $\geq2$ & $0$ & $0$ & $0$\\\hline
    \end{tabular}
    \subcaption{Symmetry class $\mathrm{C\Romannum 2}$}
    \label{Subtable:By quaternionic structures}
    \end{subtable}
    \caption{We display sets of homotopy classes $\l[\l(I^d,\p I^d\r),\l(V_{m-\abs{\nu}}\l(\C^m\r),E\r)\r]_{\Z_2}$ in the presence of canonical time-reversal symmetry realised by \subref{Subtable:By real structures} real structures or by \subref{Subtable:By quaternionic structures} quaternionic structures. The elements containing a $\star$ mean yet-to-be-evaluated sets of homotopy classes and indicate the emergence of an unstable regime for all $\abs{\nu}<\lceil d/2\rceil$.}
    \label{tab:Topological invariants with Time Reversal Symmetry} 
\end{table}

With our used tools, the topological classification in the presence of time-reversal symmetry for higher dimensions $d\geq3$ of the underlying lattice becomes more complicated compared to the lower dimensional cases. The reason is that the algorithm in \cref{eq:The Algorithm} requires the results from dimension $d$ to make progress in dimension $d+1$.

However, one peculiar connection between the Maxwell counting index and the homotopical classification is the following: In any dimension $d$ of the underlying lattice, the classification solely depends on the Maxwell counting index, that is, on the \textit{difference} between the number of spin wave degrees of freedom and ground state constraints and not on their individual values. This implies that rigidity matrices of various dimensions can represent identical strong invariants, and a necessary condition for that is that the absolute difference between their individual number of rows and columns is identical. Kane and Lubensky\cite{Kane_2013} discovered the possibility of topological protection of zero modes in mechanical systems even in the $\nu=0$ case. This is also reflected by our \Cref{Subtable:By real structures,Subtable:By quaternionic structures}. Moreover, in symmetry class $\mathrm{BDI}$, there is still the possibility of nontrivial topology for the Maxwell counting index $\abs{\nu} = 1$ for three-dimensional lattices, which might inspire future research in unfolding the entire topology of frustrated systems on higher dimensional lattices. More generally, a topological classification addressing the unstable regime is called for. That is, envisioning an enlargement of \Cref{Subtable:By real structures,Subtable:By quaternionic structures} along the horizontal dimension axis, one must imagine a diagonal running down these two tables, from top left to bottom right. This diagonal separates an unstable, that is, non-trivial, regime $\abs{\nu}<\lceil d/2\rceil$ from the topologically trivial regime $\abs{\nu}\geq\lceil d/2\rceil$. This behavior resembles the nature of homotopy groups of spheres where the Maxwell counting index $\abs{\nu}$ plays an analogous role to the dimension of a sphere. Although spheres are well-studied $\mathrm{CW}$ complexes, their homotopy groups are still under intensive research by algebraic topologists\cite{hatcher2002algebraic_stiefel}. 

In the absence of time-reversal symmetry, one classifies through the homotopy groups of complex Stiefel manifolds $\pi_d\l(V_{m-\abs{\nu}}\l(\C^m\r)\r)$. These homotopy groups are displayed in \Cref{tab:Homotopy groups of complex Stiefel manifolds} up to dimensions $d = 6$ and $\abs{\nu} = 3$ for which the first trivial row appears. 

\begin{table}[H]
\centering
   \begin{tabular}{|c|c|c|c|cc|cc|ccc|}\hline
         \multirow{3}{*}{$\abs{\nu}$} &\multicolumn{10}{|c|}{Dimension $d$}\\\cline{2-11}
            &  $1$ & $2$ & $3$ & \multicolumn{2}{c|}{$4$} & \multicolumn{2}{c|}{$5$} & \multicolumn{3}{c|}{$6$}\\
          & & & $m\geq 2$ & $m=2$ & $m\geq3$ & $m = 2$ & $m\geq3$ & $m = 2$ & $m = 3$ & $m\geq4$\\\hhline{|=|=|=|=|==|==|===|}
          $0$ & $\mathbb Z$ & $0$ &  $\mathbb Z$  & $\Z_2$ & $0$& $\Z_2$ & $\Z$ & $\Z_{12}$ & $\Z_6$ & $0$\\
          $1$ & $0$ & $0$ &  $\mathbb{Z}$  & $\Z_2$ & $0$ & $\Z_2$ & $\Z$ & $\Z_{12}$ & $\Z_6$ & $0$\\
          $2$ & $0$ & $0$ &  $0$ &  & $0$ & &$\Z$ &  & $\Z_2$ & $\Z_2$\\
          $\geq3$ & $0$ & $0$ & $0$ &  & $0$&&$0$ &  & &$0$\\\hline
    \end{tabular}
    \caption{The topological classification for symmetry class $\mathrm{A\Romannum3}$ is realized by the homotopy groups of complex Stiefel manifolds. The case $\pi_d(\U(1)) =\pi_d\l(S^1\r) = 0$ for all $d\geq2$ is omitted in this table from $d=3$. The table entries are taken from \cite{dieck2008algebraic_stiefel,gilmore_1967,Mimura_1963,hatcher2002algebraic_stiefel}.}
    \label{tab:Homotopy groups of complex Stiefel manifolds} 
\end{table}

The primary tool in computing the sets of $\Z_2$-homotopy classes portrayed in \Cref{Subtable:By real structures,Subtable:By quaternionic structures} is algorithmically illustrated in \cref{eq:The Algorithm}.
The morphisms $i_*$ and $j_*$ are the induced inclusions $i\colon \l(\Omega X\r)^{\Z_2}\hookrightarrow \Omega X$ and $j\colon\l(\Omega X,c_{x_0}\r)\hookrightarrow\l(\Omega X,(\Omega X)^{\Z_2}\r)$, respectively, and the boundary operator $\p$ is defined by evaluating representatives of homotopy classes at $-\pi$.
    \begin{equation}
     \begin{tikzcd}[row sep=1.3em, column sep=0.8em]
       & & {\l[\l(I^d,\p I^d\r),(X,x_0)\r]_{\Z_2}}\ar[d,phantom,"\nvisom"] & & &\\
     \pi_1\l(\l(\Omega^{d-1}X\r)^{\Z_2}\r)\ar[r,"i_*"]&\pi_1\l(\Omega^{d-1}X\r)\ar[r,"j_*"]&\pi_1\l(\Omega^{d-1}X,\l(\Omega^{d-1}X\r)^{\Z_2}\r)\ar[r,"\partial"] &\pi_0\l(\l(\Omega^{d-1}X\r)^{\Z_2}\r)\ar[r,"i_*"] &\pi_0\l(\Omega^{d-1}X\r) & \\
      & \pi_d(X)\ar[u,phantom,"\nvisom"]& \huge\vdots\ar[d,phantom,"\huge\vdots",yshift=1.64ex]\ar[u, phantom,"\huge\vdots", yshift=-1.43ex] & & \pi_{d-1}(X)\ar[u,phantom,"\nvisom"]&\\       \pi_1\l(\l(\Omega^2X\r)^{\Z_2}\r)\ar[r,"i_*"]&\pi_1\l(\Omega^2X\r)\ar[r,"j_*"]&\pi_1\l(\Omega^2X,\l(\Omega^2X\r)^{\Z_2}\r)\ar[r,"\partial"] &\pi_0\l(\l(\Omega^2X\r)^{\Z_2}\r)\ar[r,"i_*"] &\pi_0\l(\Omega^2X\r) & \\
      \pi_2\l(\Omega X,\l(\Omega X\r)^{\Z_2}\r)\ar[u,phantom,"\nvisom"] & \pi_3(X)\ar[u,phantom,"\nvisom"]& & & \pi_2(X)\ar[u,phantom,"\nvisom"]& \\
     \pi_1\l(\l(\Omega X\r)^{\Z_2}\r)\ar[r,"i_*"]&\pi_1\l(\Omega X\r)\ar[r,"j_*"] &\pi_1\l(\Omega X,\l(\Omega X\r)^{\Z_2}\r)\ar[uur,"\isom",controls={+(-0.0,2.6) and +(0.0,-2.6)}]\ar[r,"\partial"] &\pi_0\l(\l(\Omega X\r)^{\Z_2}\r)\ar[r,"i_*"] & \pi_0\l(\Omega X\r)& \\ 
     \pi_2\l(X,X^{\Z_2}\r)\ar[u,phantom,"\nvisom"] & \pi_2(X)\ar[u,phantom,"\nvisom"]& & &\pi_1(X)\ar[u,phantom,"\nvisom"] &\\    \pi_1\l(X^{\Z_2}\r)\ar[r,"i_*"]&\pi_1(X)\ar[r,"j_*"] &\pi_1\l(X,X^{\Z_2}\r)\ar[uur,"\isom",controls={+(-0.0,2.2) and +(0.0,-2.2)}]\ar[r,"\partial"] &\pi_0\l(X^{\Z_2}\r)\ar[r,"i_*"] & \pi_0(X)&
 \end{tikzcd}
\label{eq:The Algorithm}
\end{equation}

We stack the ends of the well-known homotopy sequences of based pairs \cite{hatcher2002algebraic} of the form $\Big(\Omega^{d-1}X,\l(\Omega^{d-1}X\r)^{\Z_2},\allowbreak c_{x_0}\Big)$ for each $d\geq1$ and connect them via the isomorphisms constructed in \Cref{homotopy groups of equivariant loop space}. We want to calculate the first relative homotopy set in the middle of each exact sequence in \cref{eq:The Algorithm}. However, the fourth set of homotopy classes in every sequence is another set of path components of a $\Z_2$-equivariant iterated loop space in one lattice dimension less that we are interested in calculating. By \Cref{homotopy groups of equivariant loop space}, this gives rise to a similar exact sequence in one lattice dimension less, and the argument iterates. We proceed with the iteration until we reach the bottom, i.e., an exact sequence in which the number of loop coordinates has been reduced to $0$. In this very sequence we determine every set or group of homotopy classes, in particular the set $\pi_0\l(\l(\Omega X\r)^{\Z_2}\r)\cong\pi_1\l(X,X^{\Z_2}\r)$, and move upwards from there by considering the next higher lattice dimension, $d = 2$, and move on.

We illustrate the application of \cref{eq:The Algorithm} by deriving \Cref{Subtable:By real structures,Subtable:By quaternionic structures}. To do so, we employ two technical preliminaries. 

\begin{lemma}[See \cite{dieck2008algebraic}]\label{First relative homotopy set as the quotient of the fundamental group modulo image of induced inclusion}
    Let $(X,A,x_0)$ be a based pair of spaces in which $A$ is path connected. Then each element in $\pi_1(X,A)$ is represented by a loop in $(X,x_0)$ and the map $j_*\colon\pi_1(X)\to\pi_1(X,A)$ induces a bijection of $\pi_1(X,A)$ with the right (or left) cosets of $\pi_1(X)$ modulo the image of $i_*\colon\pi_1(A)\to\pi_1(X)$.
\end{lemma}

\begin{figure}[H]
    \centering
    \includegraphics[width=0.6\textwidth]{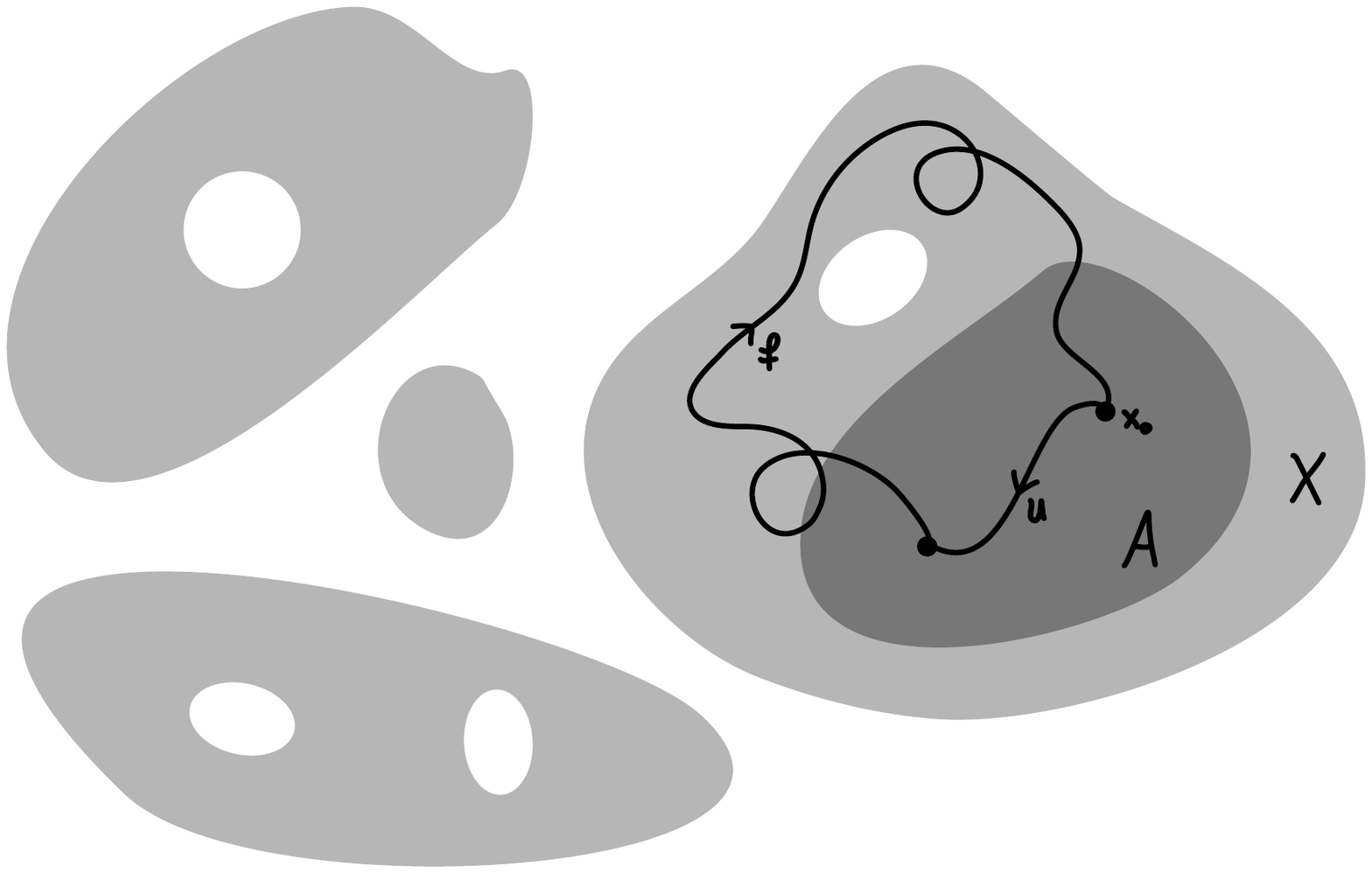}
    \caption{The main idea of \Cref{First relative homotopy set as the quotient of the fundamental group modulo image of induced inclusion}: The endpoints of a representative $f$ of the relative homotopy class $[f]\in\pi_1(X,A,x_0)$ can be connected by a path $u$ in the path connected subspace $A$ (dark grey). The concatenation $u+f$ is a loop based at $x_0$ and is homotopic to $f$, i.e. $[u+f] = [f]$. The space $X$ (light grey) must not be path connected.}
    \label{fig:Path connected subspace}
\end{figure}

Finally, we employ

\begin{figure}[H]
    \vspace{-.8cm}
    \centering
    \includegraphics[width=0.5\textwidth]{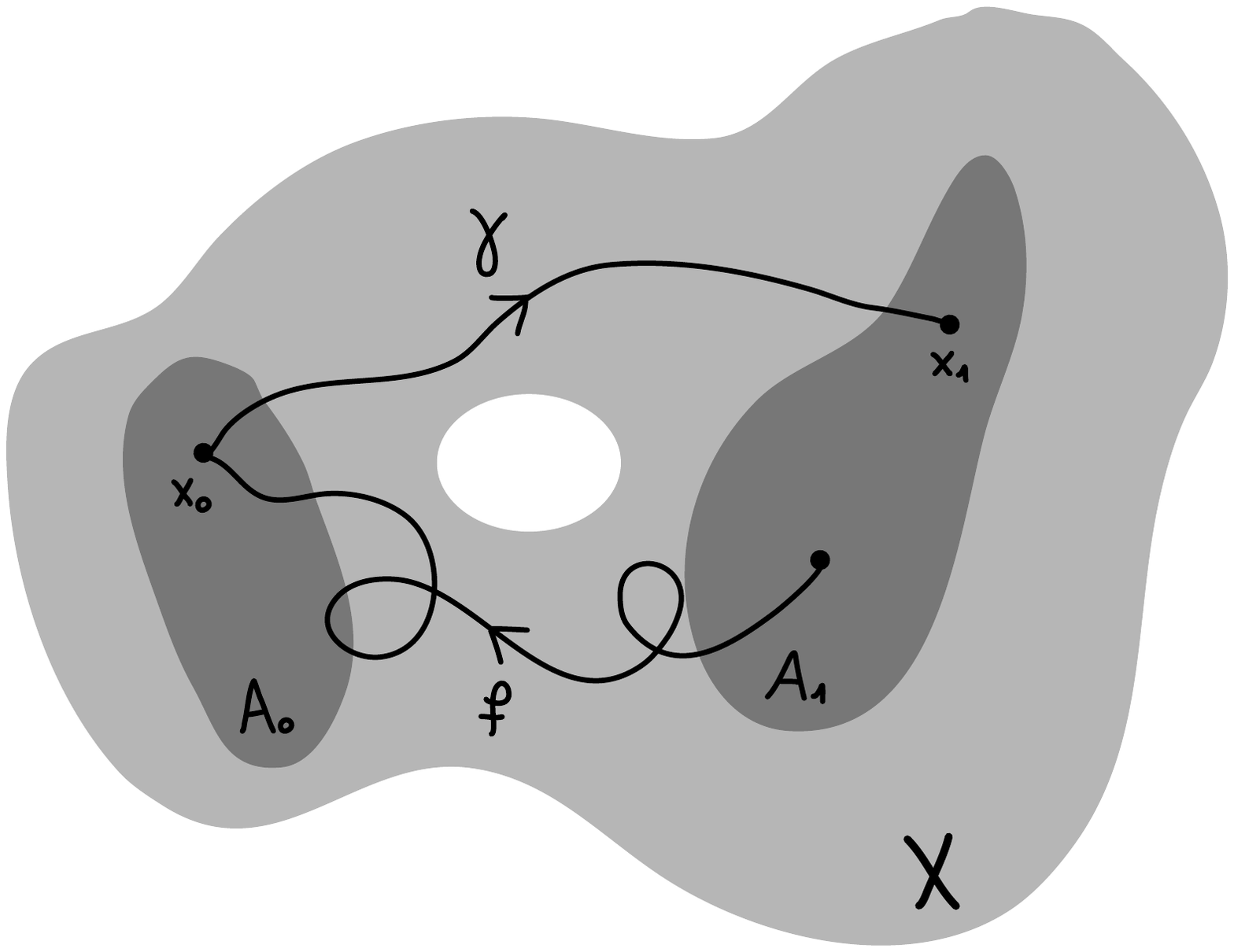}
    \caption{Depicted is a path connected space $X$ (light grey) and a subspace $A = A_0 \mathbin{\dot{\cup}} A_1$ (dark grey) consisting of two path components $A_0$ and $A_1$. Connecting the points $x_0\in A_0$ and $x_1\in A_1$ through a path $\gamma$, the concatenation $f+\gamma$ represents a relative homotopy class $[f+\gamma]\in\pi_1(X,A_1,x_1)$ for every $f\in S_1$ (see \Cref{Changing the base point induces bijection to first relative homotopy set} for the definition of $S_1$).}
    \label{fig:Bijection to first relative homotopy set}
\end{figure}

\begin{lemma}\label{Changing the base point induces bijection to first relative homotopy set}
    Let $(X,A,x_0)$ be a based pair in which $X$ is path connected and $A$ consists of countably many path components $A = \mathop{\dot{\bigcup}}\limits_{j\in J}A_j$. Choose points $x_i\in A_i$ for every $i\in J$ and consider the spaces of maps
    \begin{equation}
        S_{i}\coloneqq\l\{f\colon I\to X\mid f(-\pi)\in A_i\text{ and }f(\pi) = x_0\r\}.
    \end{equation}
    We then have a bijection $S_{i}/\sim\cong\pi_1\l(X,A_i,x_i\r)$ for every $i\in J$. Furthermore, we have a bijection \begin{equation}\pi_1(X,A,x_0)\cong\bigsqcup\limits_{j\in J}\pi_1\l(X,x_j\r)/i_*\l(\pi_1\l(A_j,x_j\r)\r).\end{equation} 
    \begin{proof}
        Let $\gamma_i\colon I \to X$ be a path from $\gamma_i(-\pi) = x_0$ to $\gamma_i(\pi) = x_i$. For $f\in S_i$ the concatenation $f+\gamma_i\colon\l(I,\p I,\{\pi\}\r)\to\l(X,A_i,x_i\r)$ defines a representative of $[f+\gamma_i]\in\pi_1(X,A_i,x_i)$ (see also \Cref{fig:Bijection to first relative homotopy set}). 

        We therefore define the map $\theta_i\colon S_i/\sim\to\pi_1(X,A_i,x_i)$ by $\theta[f]\coloneqq[f+\gamma_i]$.
        The map $\theta$ is well defined because for $[f] = [g]\in S_i/\sim$ we find a homotopy $H\colon I\times[0,1]\to X$ with $H\l(\{-\pi\}\times[0,1]\r)\subseteq A_i$ and $H\l(\{\pi\}\times[0,1]\r) = \{x_0\}$ from $H_0 = f$ to $H_1 = g$. The map $G\colon\l(I\times[0,1],\p I\times[0,1],\{\pi\}\times[0,1]\r)\to\l(X,A_i,x_i\r)$ defined by $G_t\coloneqq H_t+\gamma_i$ defines a homotopy from $G_0 = f+\gamma_i$ to $G_1 = g+\gamma_i$ implying $\theta[f] = \theta[g]$.

        Further, the map $\theta_i$ is a bijection and its inverse function $\theta_i^{-1}\colon\pi_1(X,A_i,x_i)\to S_i/\sim$ reads $\theta_i^{-1}[g] = [g+\overline{\gamma_i}]$, because $\theta_i^{-1}\theta_i[f] = [(f+\gamma_i)+\overline{\gamma_i}] = [f+(\gamma_i+\overline{\gamma_i})] = [f]$ for all $[f]\in S_i/\sim$ and $\theta_i\theta_i^{-1}[h] = [(h+\overline{\gamma_i})+\gamma_i] = [h+(\overline{\gamma_i}+\gamma_i)] = [h]$ for all $[h]\in\pi_1(X,A_i,x_i)$.

        Finally, we apply \Cref{First relative homotopy set as the quotient of the fundamental group modulo image of induced inclusion} and obtain \begin{equation*}
            \pi_1(X,A,x_0)\cong\bigsqcup_{j\in J}S_j/\sim\cong\bigsqcup_{j\in J}\pi_1\l(X,A_j,x_j\r)\cong\bigsqcup_{j\in J}\pi_1\l(X,x_j\r)/i_*\l(\pi_1\l(A_j,x_j\r)\r).
        \end{equation*}
    \end{proof}
\end{lemma}

For the case $1\leq n <m$ we use \cref{eq:The Algorithm} and obtain 
\begin{equation}\label{Diagram}
    \adjustbox{center}{
\begin{tikzcd}
    & {\l[\l(I,\p I\r),\l(V_n\l(\C^m\r),E\r)\r]_{\Z_2}}\ar[d,phantom,"\nvisom"] & \\
    \pi_1\l(V_n\l(\C^m\r)\r)\ar[r,"j_*"] & \pi_1\l(V_n\l(\C^m\r),V_n\l(\R^m\r)\r)\ar[r,"\p "] & \pi_0\l(V_n\l(\R^m\r)\r)\\
    0 \ar[u,equal] & & 0\ar[u,equal]
\end{tikzcd}}
\end{equation}
from which we deduce \begin{equation}\label{eq:Trivial regime in one dimension and by real structures}
    \l[\l(I,\p I\r),\l(V_n\l(\C^m\r),E\r)\r]_{\Z_2} = 0
\end{equation}
by exactness. In the case $n = m$, we make base points explicit and apply \Cref{Changing the base point induces bijection to first relative homotopy set} to find \begin{equation}\label{eq:Lie group case in one dimension}
\begin{aligned}
    \l[(I,\p I),\l(\U(m),I_m\r)\r]_{\Z_2}&\cong\pi_1\l(\U(m),\O(m),I_m\r)\\&\cong\pi_1\l(\U(m),\SO(m),I_m\r)\sqcup\pi_1\big(\U(m),\widetilde{\SO}(m),\tilde I_m\big)\\&\cong\Z\sqcup\Z\\&\cong\Z.
\end{aligned}
\end{equation}
Here, $\widetilde{\SO}(m)\subseteq\O(m)$ denotes the subset of orthogonal matrices with a determinant equal to $-1$ and $\tilde I_m$ denotes the diagonal matrix having one diagonal entry equal to $-1$ and every other diagonal element equal to $1$.
Explicitly, we used the bijections \begin{equation}
    \pi_1\big(\U(m),\allowbreak\widetilde{\SO}(m),\allowbreak\tilde I_m\big)\allowbreak\cong\allowbreak\pi_1(\U(m),\allowbreak\SO(m),\allowbreak I_m)\allowbreak\cong\allowbreak\pi_1(\U(m),\allowbreak I_m)/i_*\l(\pi_1(\SO(m),I_m)\r),
\end{equation} $\pi_1(\U(m))\cong\Z$ and used the fact that the based triple $(\U(m),\allowbreak\SU(m),\SO(m),I_m)$ has the following useful property. Since $\pi_1(\SU(m))=0$, the inclusion $i\colon\l(\SO(m),I_m\r)\hookrightarrow\l(\U(m),I_m\r)$ induces the trivial homomorphism $i_*\l(\pi_1\l(\SO(m),I_m\r)\r)=0$.
This is more generally true for any based triple $(X,A,B,x_0)$, inclusion $i\colon(B,x_0)\hookrightarrow(X,x_0)$ in any dimension $d\geq0$: As soon as $\pi_d(A) = 0$, representatives $f$ of any homotopy class $[f]\in\pi_d(B)$, when considered as homotopy classes $i_*[f]\in\pi_1(X)$ can be continuously deformed inside $A$ to the constant map representing the trivial homotopy class $i_*[f]=0$, and therefore $\Im\l(i_*\colon\pi_d(B)\to\pi_d(X)\r) = 0$.

With \cref{eq:Trivial regime in one dimension and by real structures} and the fact that, for $d<2(k-l)+2$\cite{fomenko},
\begin{equation}
\pi_d\l(V_l\l(\C^k\r)\r)\cong\cdots\cong\pi_d\l(V_1\l(\C^{k-l+1}\r)\r) = \pi_d\l(S^{2(k-l)+1}\r)\cong\begin{cases}
    0&\text{for }d<2(k-l)+1,\\
    \Z&\text{for }d = 2(k-l)+1,
\end{cases}
\label{eq:Homotopy groups of complex and quaternionic Stiefel manifolds}
\end{equation}
we inductively apply \cref{eq:The Algorithm} to deduce 
\begin{equation}
\label{eq:Trivial regime by real structures}
{\l[\l(I^{d},\p I^{d}\r),\l(V_n\l(\C^m\r),E\r)\r]_{\Z_2}} = 0
\end{equation} 
for all $1\leq n\leq m-\lceil d/2\rceil$. A very similar and analogous analysis holds in the quaternionic case: Here, $m$ and $n$ are necessarily even, and one must substitute $V_n\l(\R^m\r)$ in \cref{Diagram} with $V_{n/2}\l(\Q^{m/2}\r)$. Hence, we have proven \cref{trivial regime} and derived the nontrivial entry in \Cref{Subtable:By real structures}.

We now focus on \Cref{Subtable:By quaternionic structures} and derive the nontrivial entry. As usual, we employ \cref{eq:The Algorithm},

\begin{equation}\label{quaternionic sequence}
    \adjustbox{center}{
\begin{tikzcd}[column sep = small, row sep = small]
    & & {\l[\l(I,\p I\r),\l(V_n\l(\C^m\r),E\r)\r]_{\Z_2}}\ar[d,phantom,"\nvisom"]&  \\
\pi_1\l(V_{n}\l(\C^{m}\r)^{\Z_2}\r)\ar[r,"i_*"] &\pi_1\l(V_{n}\l(\C^{m}\r)\r)\ar[r,"j_*"] &\pi_1\l(V_{n}\l(\C^{m}\r),V_{n}\l(\C^{m}\r)^{\Z_2}\r)\ar[r,"\partial"]\ar[u,phantom,"\nvisom"] &\pi_0\l(V_{n}\l(\C^{m}\r)^{\Z_2}\r),\\
         \pi_1\l(V_{n/2}\l(\Q^{m/2}\r)\r)\ar[u,phantom, "\nvisom"] & & & \pi_0\l(V_{n/2}\l(\Q^{m/2}\r)\r)\ar[u,phantom,"\nvisom"] \\
         0\ar[u,equal] & & & 0\ar[u,equal]    
\end{tikzcd}}
\end{equation}
to deduce the surjectivity of the induced inclusion $j_*\colon\pi_1\l(V_n\l(\C^m\r)\r)\to\pi_1\l(V_n\l(\C^m\r),V_n\l(\C^m\r)^{\Z_2}\r)$ by the exactness of \cref{quaternionic sequence}. Moreover, since  $i_*\l(\pi_1\l(V_n\l(\C^m\r)^{\Z_2}\r)\r) = 0$, \Cref{First relative homotopy set as the quotient of the fundamental group modulo image of induced inclusion} delivers  
\begin{equation}\label{eq:Invariants for quaternionic structure in one dimension}
    {\l[\l(I,\p I\r),\l(V_n\l(\C^m\r),E\r)\r]_{\Z_2}}\cong\pi_1\l(V_n\l(\C^m\r)\r)/i_*\l(\pi_1\l(V_n\l(\C^m\r)^{\Z_2}\r)\r)\cong\begin{cases}
        0&\text{for }n<m,\\
        \Z&\text{for }n=m.
    \end{cases}
\end{equation} 
This result is similar to the first column of \Cref{Subtable:By real structures}, although the acting time-reversal operators and resulting target spaces are different.
This completes the derivation of \Cref{Subtable:By real structures,Subtable:By quaternionic structures}.

The calculation and physical interpretation of individual homotopy classes represented by rigidity matrices of a specific frustrated magnet is known without time-reversal symmetry\cite{Roy&Lawler}. For example, RL associate the nontrivial winding of the argument of the determinant of unitary matrices in the $\nu = 0$ and $d = 1$ case (see \Cref{tab:Homotopy groups of complex Stiefel manifolds})  to the existence of zero modes in the form of Weyl points. Hence, invariants, in this case, are winding numbers of curves encircling Weyl points. RL\cite{Roychowdhury_2018} go ahead to associate line nodes and Dirac strings to the topology of the case in which both time-reversal and $C_2$ rotation symmetry are present. An interesting avenue would be to calculate strong invariants in the present setting of true $\Z_2$-equivariance (see \Cref{Subtable:By real structures,Subtable:By quaternionic structures}).

We investigated the classical antiferromagnetic Heisenberg spin chain and obtained a rigidity matrix of symmetry class $\mathrm{BDI}$ that exhibits the Maxwell counting index $\nu = 0$. This case, therefore, indicates the protection of zero modes by a $\Z$ topology (see \Cref{Subtable:By real structures}).
For the HAF on the square lattice with anisotropic next-nearest neighbor exchange interactions (see \Cref{Fig:Altermagnetic state}), we investigated the cases $J_1 = 2J_i$ for both $i=2,3$ and $J_1>2J_i$ for both $i=2,3$. In the former, we obtained a rigidity matrix of symmetry class $\mathrm{BD\Romannum1}$ exhibiting a Maxwell counting index of $\nu = 0$. In the latter, we derived a rigidity matrix composed of two subsystems of symmetry class $\mathrm{BD\Romannum1}$ exhibiting the Maxwell counting index $\nu = 4$. The protection of zero modes by topology of the former case is therefore still under investigation, and the latter indicates a trivial topology associated with the Néel state in \Cref{Fig:Altermagnetic state} (see \Cref{Subtable:By real structures}). Since the rigidity matrix for the pyrochlore HAF in \cref{eq:Rigidity matrix for pyrochlore heisenberg antiferromagnet} consists of two blocks representing systems of symmetry class $\mathrm{BD\Romannum1}$ with $\nu = 2$, it indicates a trivial topology by \Cref{Subtable:By real structures}.

This topological classification is quite different from the topological classification of RL \cite{Roy&Lawler}. In addition to time-reversal symmetry, RL consider the presence of a $C_2$ rotation symmetry. Therefore, they argue that a $\Z_2$ invariant would protect the zero modes at the critical point, the unfrustrated N\'eel state would not be topology protected, and that a $\Z$ invariant would protect the frustrated state. 

To be more precise, the announced symmetry considerations of RL \cite{Roy&Lawler} are time-reversal symmetries. However, their examples hint towards the existence of an additional $C_2$ rotation symmetry, which has a similar effect on momenta in the Brillouin torus $T^d$ as canonical time-reversal symmetry does, namely $[\vb k]\mapsto[-\vb k]$. These compositions of symmetries lead to the $\Z_2$-invariance conditions $\tilde r(-\vb k)=\overline{\tilde r(\vb k)}=\tilde r(\vb k)$ (in the presence of time-reversal symmetry by real structures with an additional $C_2$-symmetry) and $\tilde r(-\vb k)=\big(I_{M/2}\otimes\sigma_2\big)\overline{\tilde r(\vb k)}\big(I_{N/2}\otimes\sigma_2\big)=\tilde r(\vb k)$ (in the presence of time-reversal symmetry by quaternionic structures with an additional $C_2$-symmetry). In particular, the fixed point condition $g\tilde r(\vb k) = \tilde r(\vb k)$ is satisfied for all momenta $[\vb k]\in T^d$. For this reason, RL classify the topology of zero modes in frustrated systems with \begin{equation}
    \pi_0\l(\Omega^dV_n\l(\C^m\r)^{\Z_2}\r)\cong\begin{cases}
        \pi_d(V_n(\C^m))&\text{no symmetries,}\\
        \pi_d(V_n(\R^m))&\text{for }T_i^2=+\Id\text{ and }C_2\text{-symmetry},\\
        \pi_d(V_{n/2}(\Q^{m/2}))&\text{for }T_i^2=-\Id\text{ and }C_2\text{-symmetry},
    \end{cases}
\end{equation}
i.e., the homotopy groups of complex, real, and quaternionic Stiefel manifolds. In our case, considering the presence of $\Z_2$-equivariance conditions portrayed in \Cref{Different equivariance conditions}, the fixed point condition is in particular satisfied for all $2^d$ time-reversal invariant momenta in $\l(T^d\r)^{\Z_2}$. We classify the topology with sets of path components of $\Z_2$-equivariant iterated loop spaces of complex Stiefel manifolds 
\begin{equation}
    \pi_0\l(\l(\Omega^dV_n\l(\C^m\r)\r)^{\Z_2}\r)\cong\pi_1\l(\Omega^{d-1}V_n\l(\C^m\r),\l(\Omega^{d-1}V_n\l(\C^m\r)\r)^{\Z_2}\r).
\end{equation}
Moreover, as the rigidity matrices computed by RL satisfy $\tilde r(-\vb k) = g\tilde r(\vb k)=\tilde r(\vb k)$ for all $[\vb k]\in T^d$, an additional avenue would be to classify zero modes in frustrated systems by 
\begin{equation}
    \pi_0\l(\l(\Omega^dV_n(\C^m)^{\Z_2}\r)^{\Z_2}\r)\cong\pi_1\l(\Omega^{d-1}V_n(\C^m)^{\Z_2},\allowbreak\l(\Omega^{d-1}V_n(\C^m)^{\Z_2}\r)^{\Z_2}\r)
\end{equation}
with the algorithm outlined in \cref{eq:The Algorithm}. This shows two things: First, to incorporate all symmetries of a frustrated system into a homotopical classification, one cannot sidestep the use of \Cref{homotopy groups of equivariant loop space}, in particular, \cref{eq:The Algorithm}, as long as true and nontrivial $\Z_2$-equivariance is present. Second, the homotopy classes in the examples of RL\cite{Roy&Lawler} include in our homotopy classes due to the inclusion of the real into the complex Stiefel manifold. 

\section{Conclusion}

In this paper, we proved a generalization of a result in $\Z_2$-equivariant homotopy theory and applied it to frustrated spin systems. The presented applications lead to a homotopical classification of frustrated systems in the presence or absence of canonical time-reversal symmetry. This homotopical classification of rigidity matrices explains the robust nature of frustration in the form of an accidental degeneracy of ground states in many frustrated magnets by relating it to topological invariants \cite{Roy&Lawler}. The zero modes in frustrated magnets are described in the framework of rigidity operators $R$ whose kernels contain the zero modes. The linearized Hamiltonian is given as a bilinear form in terms of $R$. Moreover, this homotopical classification can also be applied to frustrated $n$-vector models.

We further demonstrated the emergence of an unstable regime concerning the computation of sets of homotopy classes starting from lattice dimension $d = 2$ for $\abs{\nu}<\lceil d/2\rceil$, differentiated from the trivial regime $\abs{\nu}\geq\lceil d/2\rceil$. 
To complete the topological classification in dimension $d = 2$ and go beyond, one should understand more deeply the topological structure of each path component of $\l(\Omega^dV_n\l(\C^m\r)\r)^{\Z_2}$. 

The topological classification is exemplified through the antiferromagnetic Heisenberg spin chain, the pyrochlore HAF, the HAF on a square lattice with anisotropic next-nearest neighbor interactions, and the $J_1-J_2$ HAF on a square lattice. The results are compared to the ones of RL \cite{Roy&Lawler}, and it is observed that the homotopy classes in their examples include into our homotopy classes due to the inclusion of the real into the complex Stiefel manifold. 

As a next task, one could ask oneself how to calculate \begin{equation}
    \l[\l(T^d,[\vb 0]\r),\l(V_{m-\abs{\nu}}\l(\C^m\r),E\r)\r]_{\Z_2}
\end{equation} for all $d\geq1$ to gain access to strong and weak topological invariants. Leaving out the base point preservation condition, one would generalize to the sets of free $\Z_2$-homotopy classes\cite{dieck1987transformation,kennedy&guggenheim,hatcher2002algebraic_basepoint} \begin{equation}
    \l[T^d,V_{m-\abs{\nu}}\l(\C^m\r)\r]_{\Z_2}\cong\l[\l(T^d,[\vb 0]\r),\l(V_{m-\abs{\nu}}\l(\C^m\r),E\r)\r]_{\Z_2}/\pi_1\l(V_{m-\abs{\nu}}\l(\C^m\r)^{\Z_2}\r)
\end{equation} and \begin{equation}
    \big[S^d,\allowbreak V_{m-\abs{\nu}}(\C^m)\big]_{\Z_2}\allowbreak \cong\big[\big(I^d,\p I^d\big),\big(V_{m-\abs{\nu}}\l(\C^m\r),E\big)\big]_{\Z_2}/\pi_1\l(V_{m-\abs{\nu}}\l(\C^m\r)^{\Z_2}\r)
\end{equation} for a homotopical classification of time-reversal symmetric frustrated magnets. One could ask further whether there exists a product decomposition of the set $\l[T^d,V_{m-\abs{\nu}}\l(\C^m\r)\r]_{\Z_2}$ into factors of the already investigated sets of homotopy classes $\big[\big(I^d,\p I^d\big),\allowbreak\big(V_{m-\abs{\nu}}(\C^m)\big)\big]_{\Z_2}$. Or, examine whether there is at least an embedding of the form $\l[I^d/\p I^d,V_{m-\abs{\nu}}\l(\C^m\r)\r]_{\Z_2}\hookrightarrow\big[T^d,\allowbreak V_{m-\abs{\nu}}\l(\C^m\r)\big]_{\Z_2}$. Both of these avenues are true in the context of topological insulators and superconductors (when replacing complex Stiefels manifold with the symmetric spaces in the Bott-Kitaev periodic table) \cite{kennedy&guggenheim}.

As a technical generalization, one shall consider a disjoint union of the form $0\sqcup\bigsqcup_{p = 1}^qV_{n-q+p}(\C^m)$ as target spaces for rigidity matrices. This incorporates rigidity matrices into the classification whose singular values exhibit zeros at specific momenta in the Brillouin torus. Here, $q-1\in\{0,\ldots,n-1\}$ denotes the number of singular values that each have a zero at a certain momentum.  

Moreover, one shall investigate more symmetries, e.g., various crystalline symmetries, possibly retrieving $\Z_2$-equivariance conditions to realize a homotopical classification through \Cref{homotopy groups of equivariant loop space}. In three dimensions, there are already 230 crystallographic space group types. In chapter \ref{section:Examples} we were inspired by the altermagnetic Hubbard model \cite{das2023realizing} and derived the corresponding rigidity matrix in \cref{eq:Rigidity matrix for anisotropic nnn interactions}. One could incorporate the $\Z_2$-equivariance condition $\tilde r(-\vb k) = [\l((-\sigma_1)\oplus(\sigma_1\otimes\sigma_1)\r)\oplus\l(\sigma_1\oplus(-\sigma_1\otimes\sigma_1)\r)]\tilde r(\vb k)\l[\sigma_1\oplus\sigma_1\r]$ into the homotopical classification by calculating \begin{equation}
    \pi_0\bigg(\Big(\Omega^2\allowbreak V_2\big(\allowbreak\C^6\big)^{\Z_2}\Big)^{\Z_2}\bigg),
\end{equation} where the $\Z_2$-action on $V_2\l(\C^6\r)$ is a combined $\Z_2$-action and either realised by $gA\coloneqq [(-\sigma_1)\oplus(\sigma_1\otimes\sigma_1)]\overline A\sigma_1$ or by $gA\coloneqq [\sigma_1\oplus(-\sigma_1\otimes\sigma_1)]\overline A\sigma_1$. The $\Z_2$-action on  $\Omega^2V_2\l(\C^6\r)^{\Z_2}$ is realised by $(gf)(\vb k) = \overline{f(-\vb k)}$.

Finally, as the mean-field Hamiltonian of the altermagnetic Hubbard model \cite{das2023realizing} is momentum-inversion symmetric, one can employ \Cref{homotopy groups of equivariant loop space} to obtain a homotopical classification of such Hamiltonians through the algorithm in \cref{eq:The Algorithm}. 

\section*{Author Declarations}
\subsection*{Conflict of Interest}
The author has no conflicts to disclose.
\subsection*{Author Contributions}
\textbf{Shayan Zahedi:} Writing – original draft (equal); Writing – review \& editing (equal). 
\section*{Data Availability}
Data sharing is not applicable to this article as no new data were created or analyzed in this study.



%
%

%

\appendix
\section{On extending \Cref{Subtable:By real structures,Subtable:By quaternionic structures}}
The Lie group case in \cref{eq:Lie group case in one dimension} gives in particular $\l[(I,\p I),(S^1,1)\r]_{\Z_2}\cong\Z$. We wish to find $\l[(I^2,\p I^2),(S^1,1)\r]_{\Z_2}$. We employ \cref{eq:The Algorithm} and find the following exact sequence. 

\begin{equation}\label{Diagram_circle}
    \adjustbox{center}{
\begin{tikzcd}
    & {\l[\l(I^2,\p I^2\r),(S^1,1)\r]_{\Z_2}}\ar[d,phantom,"\nvisom"] & & \\
    \pi_1\l(\Omega S^1\r)\ar[r,"j_*"] & \pi_1\l(\Omega S^1,\l(\Omega S^1\r)^{\Z_2}\r)\ar[r,"\p "] & \pi_0\l(\l(\Omega S^1\r)^{\Z_2}\r)\ar[r,"i_*"] & \pi_0\l(\Omega S^1\r)\\
    0 \ar[u,equal] & & \Z\ar[u,phantom,"\nvisom"] & \Z\ar[u,phantom,"\nvisom"]
\end{tikzcd}}
\end{equation}

If we knew that, for example, the cardinality of $\l[(I^2,\p I^2),(S^1,1)\r]_{\Z_2}$ was countably infinite, we could make the following inductive argument. Suppose there exists some $d\geq2$ for which $\pi_0\l(\l(\Omega^dS^1\r)^{\Z_2}\r)\cong\l[\l(I^d,\p I^d\r),\l(S^1,1\r)\r]_{\Z_2}\cong\Z$. In other words, the space of classifying $\Z_2$-maps $\l(\Omega^dS^1\r)^{\Z_2}$ decomposes into a countable disjoint union of path components $\l(\Omega^dS^1\r)^{\Z_2} = \mathop{\dot{\bigcup}}\limits_{i\in\Z}A_i$. We choose base points $\xi_i\in A_i$ for every $i\in\Z$ and apply \Cref{Changing the base point induces bijection to first relative homotopy set} to obtain 
\begin{equation}\label{eq:Special Inductive Argument}
\begin{aligned}
    \l[\l(I^{d+1},\p I^{d+1}\r),\l(S^1,1\r)\r]_{\Z_2}&\cong\pi_1\l(\Omega^dS^1,\l(\Omega^dS^1\r)^{\Z_2},c_1\r)\\&\cong\bigsqcup_{i\in\Z}\pi_1\l(\Omega^dS^1,\xi_i\r)/i_*\l(\pi_1\l(A_i,\xi_i\r)\r) \\&= \bigsqcup_{i\in\Z}0\cong\Z
\end{aligned}
\end{equation}
since $\pi_1\l(\Omega^dS^1,\xi_i\r)\cong\pi_{d+1}\l(S^1\r) = 0$. This would prove 
\begin{equation}
    \l[\l(I^{\tilde d},\p I^{\tilde d}\r),\l(S^1,1\r)\r]_{\Z_2}\cong\Z
\end{equation}
for all $\tilde d\geq d$.

Both, for the real and quaternionic case, using \cref{eq:The Algorithm} and the known entries of \Cref{Subtable:By real structures,Subtable:By quaternionic structures}, we find that $\l[\l(I^2,\p I^2\r),\l(\U(m),I_m\r)\r]_{\Z_2}$ fits into the following exact sequence
\begin{equation}\label{Diagram_unitary}
    \adjustbox{center}{
\begin{tikzcd}
    & {\l[\l(I^2,\p I^2\r),\l(\U(m),I_m\r)\r]_{\Z_2}}\ar[d,phantom,"\nvisom"] & & \\
    0\ar[r] & \pi_1\l(\Omega\U(m),\l(\Omega\U(m)\r)^{\Z_2}\r)\ar[r] & \Z\ar[r] & \Z.  
\end{tikzcd}}
\end{equation}

\bibliography{References}

\end{document}